\documentclass[traditabstract]{aa}
\usepackage[varg]{txfonts}
\usepackage{natbib}
\usepackage{graphicx}
\usepackage{color}
\bibpunct{(}{)}{;}{a}{}{,}
\definecolor{grey}{rgb}{0.5,0.6,0.7}
\definecolor{green}{rgb}{0.,0.6,0.}

\newcommand{\blueit}[1]{\textcolor{blue}{\it #1}}
\renewcommand{\d}{{\rm d}}
\begin{document}

\title{The universal distribution of halo interlopers \\
in projected phase space}
\subtitle{Bias in galaxy cluster concentration and velocity anisotropy?}
\author{Gary A. Mamon\inst{1,2}
\and
Andrea Biviano\inst{3}
\and
Giuseppe Murante\inst{4}
}
\institute{Institut d'Astrophysique de Paris (UMR 7095: CNRS \& UPMC), 98 bis
  Bd. Arago, F--75014 Paris, France
\and
Astrophysics \& BIPAC, University of Oxford, Keble Rd, Oxford OX13RH, UK
\and
INAF, Osservatorio Astronomico di Trieste, Trieste, Italy
\and 
INAF, Osservatorio Astronomico di Torino, Torino, Italy
}
\date{Received \emph{23 December 2009} / Accepted \emph{1 July 2010}}
\abstract{
When clusters of galaxies are viewed in projection, one cannot avoid picking
up a fraction of foreground/background interlopers, that lie within the
virial cone, but outside the virial sphere.  Structural and kinematic
deprojection equations are known for the academic case of a static Universe,
but not for the real case of an expanding Universe, where the Hubble flow (HF)
stretches the line-of-sight distribution of velocities.
Using 93 mock relaxed clusters, built from the dark matter (DM) particles of a
hydrodynamical cosmological simulation, we quantify the distribution of
interlopers in projected phase space (PPS), as well as the biases in the
radial and kinematical structure of clusters produced by
the HF.
{The stacked mock clusters are well fit by an $m$=5 Einasto 
DM density profile (but only out to 1.5 virial radii), with velocity anisotropy
(VA) close to the  Mamon-{\L}okas 
model with characteristic radius equal to that of density slope $-2$.
The surface density of interlopers is nearly flat out to the virial radius, 
while their velocity
distribution shows a dominant gaussian cluster-outskirts
component and a
 flat field component.  
This distribution of interlopers in PPS is nearly universal,
showing only small trends with cluster mass, and is quantified.
A local $\kappa$=2.7 sigma velocity cut
is found to return the line-of-sight velocity dispersion profile (LOSVDP) expected
from the NFW density and VA profiles measured in three
dimensions. The HF causes a shallower outer LOSVDP that cannot be
well matched by the Einasto model for any value of $\kappa$.
After this velocity cut, which
removes 1 interloper out of 6, interlopers still account for 23$\pm1$\% of all
DM particles with projected radii within the virial radius (surprisingly very similar to the
observed fraction of cluster galaxies lying off the Red Sequence) and over
60\% between 0.8 and 1 virial radius.
The HF causes the best-fit projected NFW or $m$=5 Einasto model to the
stacked cluster to underestimate the true concentration measured in 3D by 
6$\pm$6\% (16$\pm$7\%) after (before) the velocity cut. These biases in
concentration are reduced by over a factor two once a constant
background is included in the fit.
The VA profile recovered from the measured LOSVDP 
by assuming the correct mass
profile recovers fairly well the VA measured in 3D, with a slight,
marginally significant,
bias towards more radial orbits in the outer regions.
}
{These small biases
in the concentration and VA of the galaxy system are overshadowed by important cluster-to-cluster 
fluctuations caused by cosmic variance and by the strong inefficiency
caused by the limited numbers of observed galaxies in clusters.}
An appendix provides an analytical approximation to the surface density,
projected mass and tangential shear profiles of
the Einasto model. Another derives the expressions for the surface density
and mass profiles of the NFW model
projected on the sphere (for future kinematic modeling).
}


\titlerunning{Universal distribution of halo interlopers}
\authorrunning{G. A. Mamon, A. Biviano \& G. Murante}

\maketitle

\section{Introduction}
\label{intro}
The galaxy number density profiles of
groups and clusters of 
galaxies falls off slowly enough at large radii that material beyond the
virial radius (within which these structures are thought to be in dynamical
equilibrium) contribute non-negligibly to the projected view of cluster,
i.e. to the radial profiles of surface density, line-of-sight velocity
dispersion and higher velocity moments.

In principle, this contamination of observables by \emph{interlopers},
defined here as particles that lie within the virial cone but outside the
virial sphere, is not a 
problem, since we know 
how to express deprojection equations when interlopers extend to infinity
along the line-of-sight. Consider the projection equation
\begin{equation}
\Sigma(R) = \int_{-\infty}^{+\infty} \nu(r)\, \d s = 2\,\int_R^\infty
\nu(r) {r\,\d r\over \sqrt{r^2-R^2}} \ ,
\label{projec}
\end{equation}
where $\Sigma$ and $\nu$ are the projected and space number densities,
respectively, while $R$ and $r$ are the projected and space radial distances
(hereafter, radii), respectively.
Equation~(\ref{projec}) 
can be deprojected through Abel inversion\footnote{Alternatively, the
  projection equation~(\ref{projec}) corresponds to a convolution and can
  therefore be deprojected with Fourier methods (see Discussion in
    \citealp{MB10} and references therein).} 
 to yield 
\begin{equation}
\nu(r) = -{1\over \pi}\int_r^\infty {\d \Sigma/\d R \over \sqrt{R^2-r^2}}\,\d R
\ .
\label{deprojec}
\end{equation}
The projection to infinity is explicit in equations~(\ref{projec}) and
(\ref{deprojec}).

However, the Hubble expansion
complicates the picture, as the Hubble flow moves background
(foreground) objects to high positive (negative) line-of-sight velocities.
\begin{figure}[ht]
\includegraphics[width=\hsize]{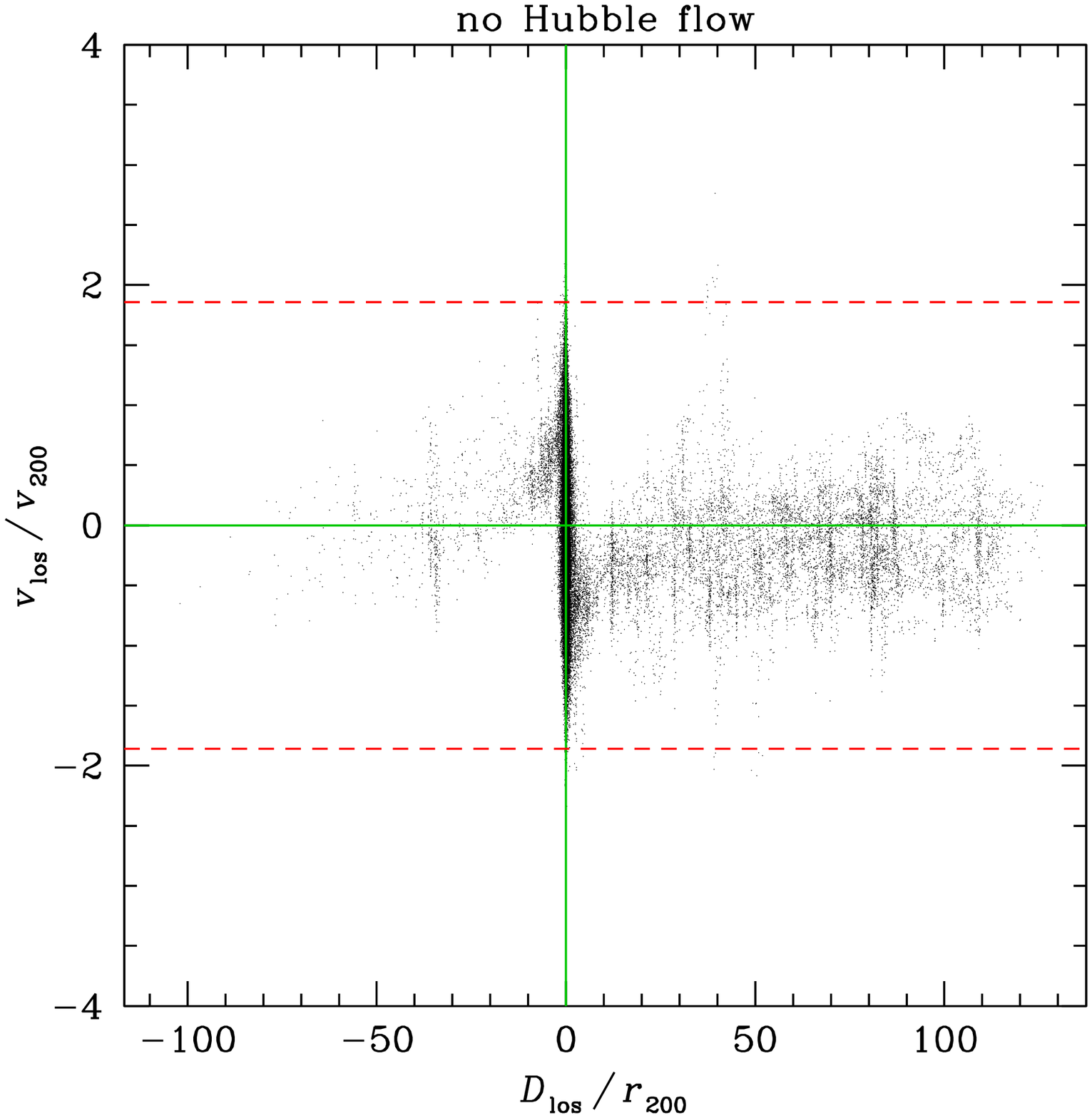}
\includegraphics[width=\hsize]{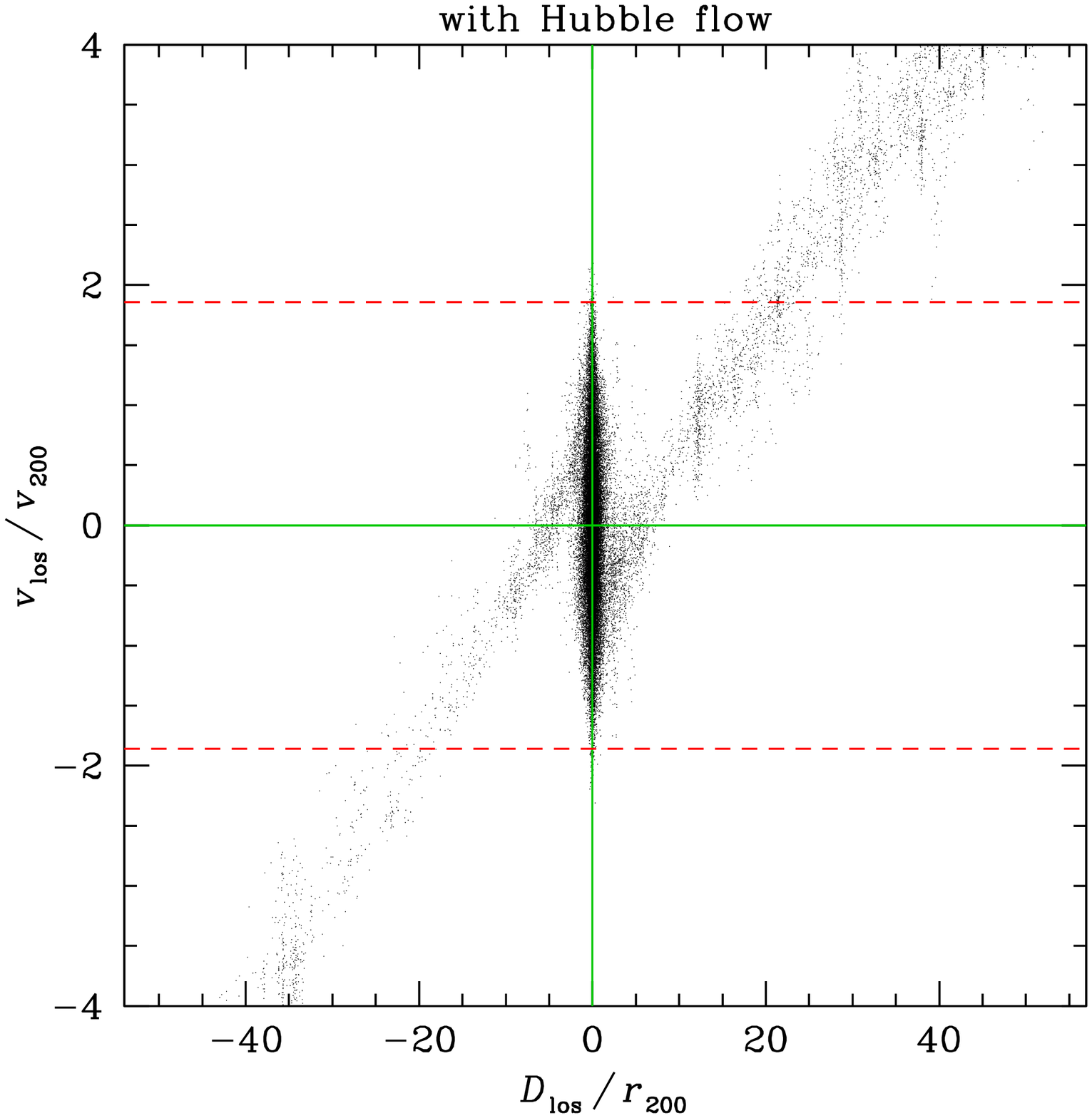}
\caption{Line-of-sight velocity as a function of real-space line-of-sight
  distance (see 
  Fig.~\ref{scheme}) for particles inside the virial cone
obtained by stacking 93
  cluster-mass halos in the cosmological simulation described in
  Sect.~\ref{data} 
without (\emph{top}) and with (\emph{bottom}) the Hubble flow (1 particle in
5 is shown for clarity).
The \emph{red dashed horizontal lines}  roughly indicate the effects of a
radius-independent $3\,\sigma$ clipping.
Note that the velocity-distance relation without Hubble flow shown here is
not entirely realistic, because the simulation was run in the context of an
expanding Universe (and cannot be run in a static Universe, for lack of
knowledge of realistic initial conditions), but should be accurate enough to
illustrate our point.
}
\label{hubflow}
\end{figure}
This is illustrated in Figure~\ref{hubflow} which shows
how the line-of-sight
velocity vs. real-space distance relation is affected by the Hubble flow.
The line-of-sight distances are computed as the segment length QP in
Figure~\ref{scheme}. 
\begin{figure}[ht]
\includegraphics[height=\hsize,angle=-90]{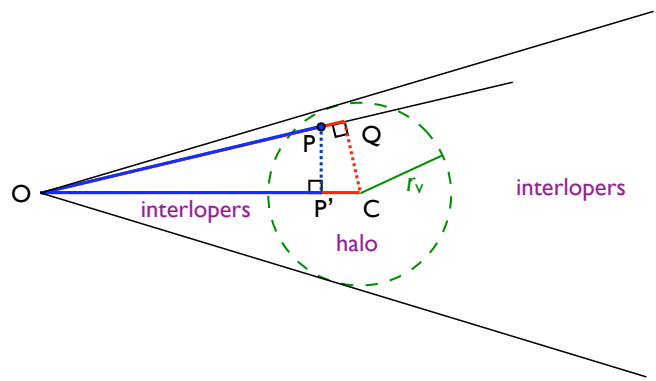}
\caption{Representation of the virial cone with halo particles inside the
  inscribed virial sphere and interlopers outside. Also shown is our
  definition of projected radius (CQ) and line-of-sight distance (OP and QP
  respectively in the observer and halo reference frames) for a random point
  P. For illustrative purposes, the distance to
  the cone is taken to be very small, so that the cone opening
  angle is much larger than in reality.
\label{scheme}}
\end{figure}
Now,
clipping the velocity differences to, say, $\kappa = 3$ times the cluster velocity
dispersion (averaged over a circular aperture, hereafter aperture velocity
dispersion), $\sigma_v$, 
gets rid of all the distant interlopers.
More precisely, the radius, $r_{\rm max}$, where the Hubble flow matches
$\kappa\,\sigma_v$ is found by solving $H_0\,r_{\rm max} = \kappa\,\sigma_v$
(where $H_0$ is the Hubble constant)
yielding 
\begin{equation}
{r_{\rm max} \over r_{\rm v}} = \kappa\,\sqrt{\Delta\over 2}\,\left ({\sigma_v\over v_{\rm v}}\right )
\ ,
\label{rmax}
\end{equation}
where $r_{\rm v}$ is the virial radius where the mean density is $\Delta$ times
the critical density of the Universe, $\rho_{\rm c} = 3 H_0^2/(8\pi G)$
(where $G$ is the gravitational constant), and
where $v_{\rm v} = \sqrt{\Delta/2}\, H_0\,r_{\rm v}$ is the 
circular velocity at the virial radius. Clusters are thought to have density
profiles consistent with the \citeauthor*{NFW96} profile (\citeyear{NFW96},
hereafter NFW)
profiles, 
\begin{equation}
\nu(r) = {1/\left(\ln2-1/2\right)\,\over \left(r/r_{-2}\right)\,\left(r/r_{-2}+1\right)^2}
\,\left[{M\left(r_{-2}\right)\over 4\,\pi\,r_{-2}^3}\right] \ ,
\label{rhoNFW}
\end{equation}
where $r_{-2}$ is the radius of density slope $-2$,
in number \citep*{LMS04}, luminosity \citep{LM03} and mass
(\citeauthor{LM03}; \citealp*{BG03, KBM04}), with a concentration,
$c=r_{\rm v}/r_{-2}$, of 3 to 5.
For isotropic NFW models, the aperture
velocity dispersion is (Appendix A of \citealp{MM07})
$\sigma_v = \eta \,v_{\rm v}$, where $\eta\simeq 0.62$ (weakly dependent on
concentration),
and equation~(\ref{rmax}) then becomes
\begin{equation}
{r_{\rm max}\over r_{\rm v}} \simeq 
13.2\,\left ({\kappa\over 3}\right )\,\sqrt{\Delta\over 100} \ .
\label{rmax2}
\end{equation}
Equation~(\ref{rmax2}) indicates that a 3-sigma clipping will remove all
material beyond 13 ($\Delta = 100$) or 19 ($\Delta = 200$) virial
radii.\footnote{With our chosen cosmology, the overdensity at the virial
  radius is $\Delta \simeq 100$, but many authors prefer to work with
  $\Delta=200$, and we will do so too.}
So, in the deprojection equation~(\ref{deprojec}), the upper integration
limit must be set to this value of $r_{\rm max}$.
Although this effective cutoff  in line-of-sight distances is quite far
removed from the cluster, it is not clear whether there may still be a
measurable bias in the concentration of clusters that one measures by
comparing the surface density distribution of galaxies in clusters with NFW
models projected out to infinity. Moreover, it is not clear how accurate are such
measures given the finite number of galaxies observed within clusters.

Finally, it is not clear how the stretching of the velocities
affects the kinematic analyses of clusters, especially in the case of nearby
clusters where the opening angle of the cone is non-negligible, leading to an
asymmetry between the foreground and background absolute velocity
distributions. For example, is the anisotropy of the 3D velocity distribution
(hereafter velocity anisotropy or simply anisotropy)
\begin{equation}
\beta(r) = 1 - {1\over 2}\,{\left\langle v_\theta^2(r)+v_\phi^2(r)\right\rangle \over
  \left\langle v_r^2(r)\right\rangle}
\label{betadef}
\end{equation}
or equivalently\footnote{$\beta(r)$ enters the Jeans equation of local
  hydrostatic equilibrium, while ${\cal A}(r)$ is a more physical definition
  of velocity anisotropy.}
\begin{equation}
{\cal A}(r) = [1-\beta(r)]^{-1/2} = \left [{2\,\left\langle
    v_r^2(r)\right\rangle \over \left\langle
    v_\theta^2(r)+v_\phi^2(r)\right\rangle} \right]^{1/2}
\label{betapdef}
\end{equation}
affected by the Hubble flow?
One can add interlopers
beyond the virial radius as a separate component to the kinematical modeling
\citep{vanderMarel+00,Wojtak+07}.
Unfortunately, we have no knowledge of the distribution of interlopers in
projected phase space (projected distance to the halo center and
line-of-sight velocity).

This paper provides the distribution of interlopers in projected phase space
(projected distance to the halo center and line-of-sight velocity) as
measured on nearly 100 stacked halos from a well-resolved cosmological
simulation.
We additionally measure the bias in the measured surface density and
line-of-sight velocity dispersion and kurtosis 
profiles compared to those obtained in a
Universe 
with no Hubble flow, and estimate how this bias affects the recovered
concentration and velocity anisotropy of the cluster.
In this paper, we use interchangeably the terms `clusters' and `halos'.

We present in Sect.~\ref{data} the cosmological simulations we use and how
the individual halos were built. In Sect.~\ref{stack} we explain how we
stack these halos. In Sect.~\ref{stats}, we present the statistics on the
halo members and the interlopers in projected phase space. Then in
Sect.~\ref{removal}, we explain how we remove the outer interlopers and show
analogous statistics on the cleaned stacked halo in Sect.~\ref{statswohivilop}.
We proceed in Sect.~\ref{conc} 
to measure the biases induced by the Hubble flow and the 
imperfect interloper removal on the estimated
concentration parameter and anisotropy profile.
We discuss our results in Sect.~\ref{discuss}.


\section{Data from cosmological $N$-body simulations}
\label{data}
The halos analyzed in this paper were extracted
by \cite{Borgani+04}
from their large cosmological hydrodynamical simulation 
performed using the 
parallel Tree+SPH {\small GADGET--2} code \citep{Springel05}.
The simulation assumes a cosmological model with present day parameters
$\Omega_{\rm m}=0.3$, $\Omega_\Lambda=0.7$, $\Omega_{\rm
b}=0.039$, $h=H_0/(100 \, \rm km \, s^{-1}\, Mpc^{-1}) = 0.7$, and
$\sigma_8=0.8$. The box size is $L=192\, 
h^{-1}$ Mpc. The simulation used $480^3$ dark matter particles and
(initially) as many gas 
particles, for a dark matter particle mass of $4.62 \times 10^9\,
h^{-1} M_\odot$.  The softening length was set to
$22.5 \, h^{-1} \, \rm comoving\ kpc$ until $z=2$ and fixed afterwards (i.e.,
$7.5\, h^{-1}$ kpc). The simulation code includes
explicit energy and entropy conservation, radiative cooling, a uniform
time-dependent UV background \citep{HM96}, the
self-regulated hybrid multi-phase model for star formation \citep{SH03}, 
and a phenomenological model for galactic
winds powered by Type-II supernovae.

Dark matter halos were identified by \citeauthor{Borgani+04} 
at redshift $z=0$ by applying a standard
Friends-of-friends (FoF) analysis to the dark matter particle set, with linking
length 0.15 times the mean inter-particle
distance. After the FoF identification,  the center of the halo was set to
the position of its most bound particle.
A spherical
overdensity criterion was then applied 
to determine the virial radius, $r_{\rm v}=r_{200}$ of each halo.
In this manner, 117 halos were identified
within the
simulated volume, among which 105 form a complete subsample
with 
virial mass $M_{200}$ larger than $10^{14}\,h^{-1} M_\odot$, thus
representing a sample of mock galaxy clusters.
Their mean and maximum masses are respectively 
$2.0 \times 10^{14}\,h^{-1} M_\odot$ and $1.1 \times 10^{15}\,h^{-1}
M_\odot$.

To save computing time, we worked on a random subsample of 
roughly 2 million
particles among the $480^3$.
Although the simulation also produced galaxies, we chose to use the dark
matter particles as tracers of the galaxy distribution for two reasons: 
1) Simulated galaxy properties in cosmological simulations depend
on details of the baryon
physics implemented in the code, and can show some mismatch with
observed properties (e.g. \citealp{Saro+06});
2) only a handful of simulated clusters had over 50 galaxies
(\citeauthor{Saro+06}), so we would have
strongly suffered from small-number statistics.
There is some debate on whether the velocity distribution of galaxies is
biased relative to the dark matter.
On one hand, the galaxy velocity distribution, although close to the dark
matter one, shows a preference for lower velocities \citep{Biviano+06},
perhaps as a consequence of dynamical friction. On the other hand, the
velocity distribution of subhalos, selected with a minimum mass before
entering their parent cluster-mass halos, is similar to that of the dark matter
\citep{FD06}. 

We visually inspected each of the 105 clusters in redshift space 
along three orthogonal viewing
axes, and removed 12 clusters that appeared, within
$r_{200}$,  to be composed of two or three
sub-clusters of similar mass (where the secondary had at least 40\% of the
mass of the primary). Most observers would omit such clusters when analyzing
their radial structure or internal kinematics.
This leaves us with 93 final mock clusters.\footnote{Including
  all 105 halos makes virtually no difference for the results of this
  article.} 
The median values (interquartile uncertainties) of their 
virial radii, virial masses, virial circular velocities,
and velocity dispersions (within
their virial spheres) are
respectively 
864$\pm$$81\,h^{-1} \, \rm kpc$,
1.50$\pm$$0.44\,10^{14}\, h^{-1}M_\odot$,
865$\pm$$81 \, \rm km \, s^{-1}$, and
584$\pm$$60 \, \rm km \, s^{-1}$.

\section{Stacking the virial cones}
\label{stack}
For each cluster, we projected the coordinates along the virial cone
(circumscribing the virial sphere, see Fig.~\ref{scheme}), as follows.
We first renormalized the 6 coordinates of phase space of the
entire simulation box to be relative to the cluster. So the cluster
most bound particle should be at the origin and its mean peculiar (bulk) velocity
should be zero.\footnote{Observers usually adopt the position of the
  brightest cluster galaxy as the center, and this corresponds to the most
  bound galaxy, so they should not suffer from important centering errors,
  although admittedly some clusters like Coma have two brightest galaxies.}
To take into account the
periodic boundaries of the simulation box, we added or
subtracted a box length to those particles situated at over a half-box length
from the cluster center. In this fashion, each cluster now effectively sits
at the center of the simulation box. We then placed an observer at
coordinates $(-D,0,0)$, $(0,-D,0)$, or $(0,0,-D)$, with $0 < D < L/2$.
We present here the results for $D=90 \,
h^{-1} \, \rm Mpc$, corresponding to a typical distance of observed clusters
in the local Universe. 
At this distance, the median virial angular radius of our 93 clusters is
33$\pm$$3\,\rm arcmin$.
We do not expect that the results of this paper should depend
on the adopted value of $D$.
We assume that the observer's peculiar velocity is equal to the cluster's
bulk velocity, so that the observer's
velocity is zero in the renormalized coordinate system.\footnote{In our
  simulation, the one-dimensional cluster bulk velocity dispersion is  $228
  \, \rm km \, s^{-1}$, so given our adopted distance of $D=90 \, h^{-1} \,
  \rm Mpc$, the typical cluster bulk velocity is only $228/9000 =
  2.5\%$. Therefore, our neglect of the observer's peculiar motion relative
  to the cluster is an adequate assumption.}

We then measured, for each cluster and for each of these three observers, the coordinates of all
2 million particles in
both the observer frame and the cluster frame.
Given the distance $r_{\rm o}$ of the particle to the observer and the projected
coordinate of the particle in cylindrical coordinates $R_{\rm c}$, 
we determined the projected distances in the observer frame (measured in a
plane perpendicular to the line-of-sight passing through the cluster center) as
$R_{\rm o}=D \,R_{\rm c}/r_{\rm o}$ (see Fig~\ref{scheme}, where $D = \rm
OC$, $r_{\rm O} = \rm OP$, $R_{\rm c} = \rm PP'$, and $R_{\rm o} = \rm CQ$).
This projection ensures that particles along the surface of the virial cone
have $R=r_{\rm v}$.
We were then able to select
all particles within a cone circumscribing the  sphere of
radius $r_{\rm v}$, where we chose $r_{\rm v} = r_{200}$, as well as $r_{\rm
  v} = 1.35\,r_{200} \simeq r_{100}$. 
In practice, we extracted data from a wider cone, circumscribing the sphere
of radius $3\,r_{200}$.

We next added the Hubble flow (for both the observer and cluster frames),
using
$H_0 = 100 \,\rm km \,s^{-1} \, Mpc^{-1}$.\footnote{All our positions and
cluster virial
radii were expressed in $h^{-1}\,\rm kpc$; the choice of $H_0$ does not matter
as long as we normalize to the virial quantities $r_{\rm v}$ and $v_{\rm v}$, which we
computed with the same 
value of $H_0$.}
We then limited in depth to line-of-sight
velocities within $4\,v_{\rm v}$ from the cluster. 
As mentioned in Sect.~\ref{intro}, the Hubble flow
effectively limits the depth of the cones to a half-length of
$\hat\kappa\,\sqrt{\Delta/2} = 40$ virial radii (see eq.~[\ref{rmax}]), 
where the cut in velocity space
in units of virial velocity is $\hat\kappa = 4$ and the
virial overdensity is $\Delta =
200$.\footnote{Throughout this paper, we use $\hat x$ to express quantity $x$ in
  virial units.}
Our results should not depend much on the distance of our observer ($90/0.864
= 104$ virial radii), as our first cut at $\pm$4 virial velocities limits the
line of sight to 40\% of the observer's distance.

We finally normalized the particle
relative positions and velocities to the virial radius, $r_{\rm v}$ 
and circular
velocity at the virial radius, $v_{\rm v}$, respectively, and finally computed the
projections of the velocities in spherical coordinates (to later measure
the 3D radial profiles of density and velocity anisotropy). For clarity, we
will sometimes use the notation 
$r_{200}$ for the virial radius and $v_{200}$ for the virial velocity.

In the end, we have roughly 84 thousand particles for each of the three
\emph{cartesian} stacked virial cones, which we then stack together into our
\emph{global} stacked
virial cone, with a grand total of
roughly a
quarter-million
particles,
among which nearly three-quarters 
lie within the virial radius. Hence, roughly 1/30th 
of all the particles in the simulation box lie
within the virial radius, $r_{200}$, of our 93 clusters.
Note that with
our stacking method, some of the particles inside virial
cones but
outside virial radii can end up being
selected in more than one of the three cartesian stacked virial cones.
However, this fraction is small (27\% of the interlopers, 
i.e. less than 8\% of all particles in virial cones),
so the three cartesian stacked virial cones are virtually independent (except
that their halos are common), hence can be
stacked into the global  virial cone.

\begin{figure}[ht]
\includegraphics[width=\hsize]{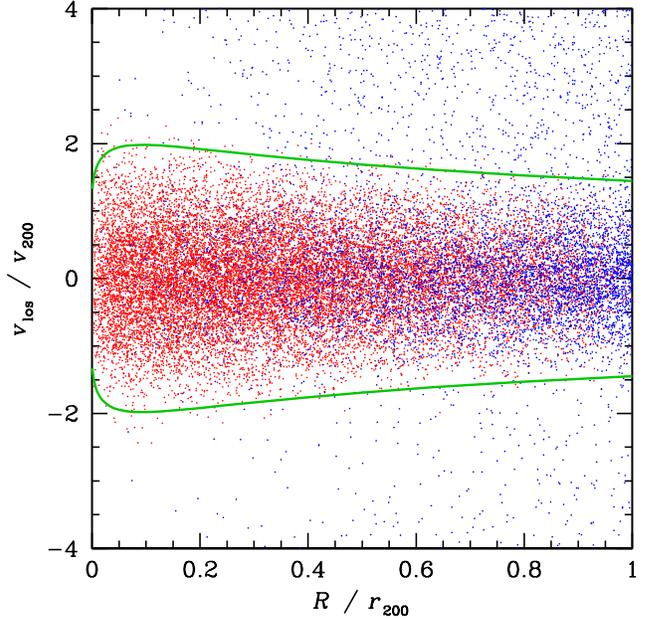}
\caption{Projected phase space diagram of stacked virial cone (built from
  $3\times93$ halos). 
Only 1 particle in
  5 is shown for clarity. The \emph{red (lighter)} and \emph{blue (darker)} points
  refer to the 
  particles within and outside the virial sphere, respectively.
The   \emph{green curves} illustrate the $\pm$$2.7\,\sigma_{\rm los}(R)$
  velocity cut  (from
eqs.~[\ref{siglosapx}] and [\ref{siglosconvert}])
for the $c\equiv c_{200} =4$ NFW model with $r_{\rm a}$=$r_{-2}$ ML anisotropy
(eq.~[\ref{betaML}]).
\label{phasespacewcuts}}
\end{figure}
Figure~\ref{phasespacewcuts} shows the projected phase space (line-of-sight
velocity, $v_{\rm los}$, versus projected radius, $R$) distribution of
particles of the stacked cluster, highlighting the 
\emph{halo particles} (lying within the virial sphere) and 
\emph{interlopers} (lying in the virial cone but outside of
the virial sphere).
Note that the halo particles are confined to fairly
small velocities (the largest absolute halo particle velocity is
$2.7\,v_{\rm v}$).
One notices an excess of positive velocity outliers in comparison with
negative velocity ones, as expected from the conical projection used here.

\section{Interloper statistics before the velocity cut}
\label{stats}

We now measure the distribution of interlopers in projected phase space and
study its dependence on halo mass.

\subsection{Global statistics}

\begin{figure}[ht]
\centering
\includegraphics[width=\hsize]{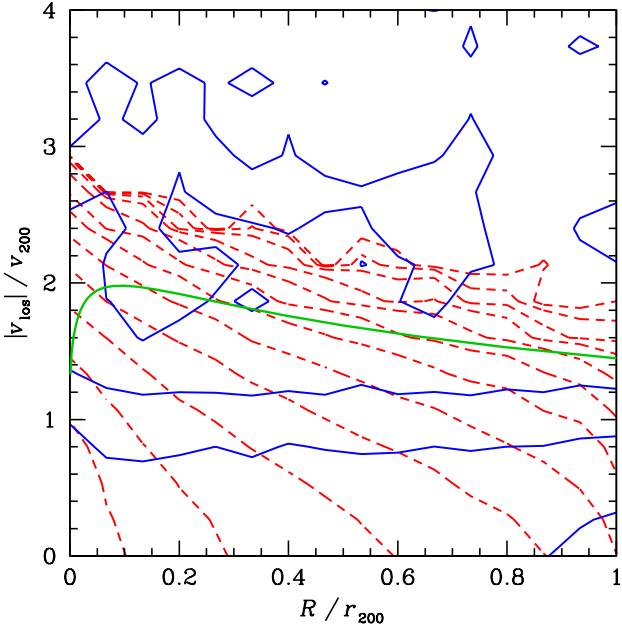}
\caption{Contours of projected phase space density of stacked virial cone ---
 in units of ${\rm d}^2N/(\d R/r_{200}\,\d |v_{\rm los}|/v_{200})/(2\pi R/r_{200})$ ---
 of halo (\emph{dashed red
    contours}) and interloper (\emph{solid blue}) particles. The halo contours are
    equally spaced in log-density, increasing by a factor 2.44 from 0.085 (upper
    right) to 3370 (lower left). The interloper contours are taken from the
same    set as the halo contours, but
    limited from the 5th highest level halo contour (72 virial units,
  at the lower left) to the 8th highest level halo contour (4.0 virial units,
  for all contours above the 3rd nearly horizontal one).
The   \emph{green curve} shows the $2.7\,\sigma_{\rm los}(R)$ (from
eqs.~[\ref{siglosapx}] and [\ref{siglosconvert}])
  velocity cut for the $c$=4 NFW model with $r_{\rm a}$=$r_{-2}$ ML anisotropy
(eq.~[\ref{betaML}]).
}
\label{Rvzdist}
\end{figure}
Figure~\ref{Rvzdist} displays contours of the density in $(R,|v_{\rm los}|)$ projected phase
space. Note that the projected phase space density of
Figure~\ref{Rvzdist} is proportional to ${\rm d}^2 N /(R \,\d R \,\d |v_{\rm los}|)$, hence the
different shapes than seen in Figure~\ref{phasespacewcuts}.
The interlopers have a very different projected phase space density than the halo
particles.
In particular, their horizontal contours mean that 
the interloper projected phase space
density is fairly independent of projected radius.
Moreover, the interloper contours do not extend beyond $v_{\rm los} = 2\,v_{\rm v}$, except
for a few islands (caused by cosmic variance),
indicating that the velocity distribution of interlopers is close to flat at
large velocities (the islands thus represent small, probably statistical,
fluctuations in a flat background).

These issues can be looked in more detail
through slices of the projected phase space density in velocity and
radial space. 
\begin{figure}[ht]
\centering
\includegraphics[width=0.87\hsize,bb=55 10 450 745]{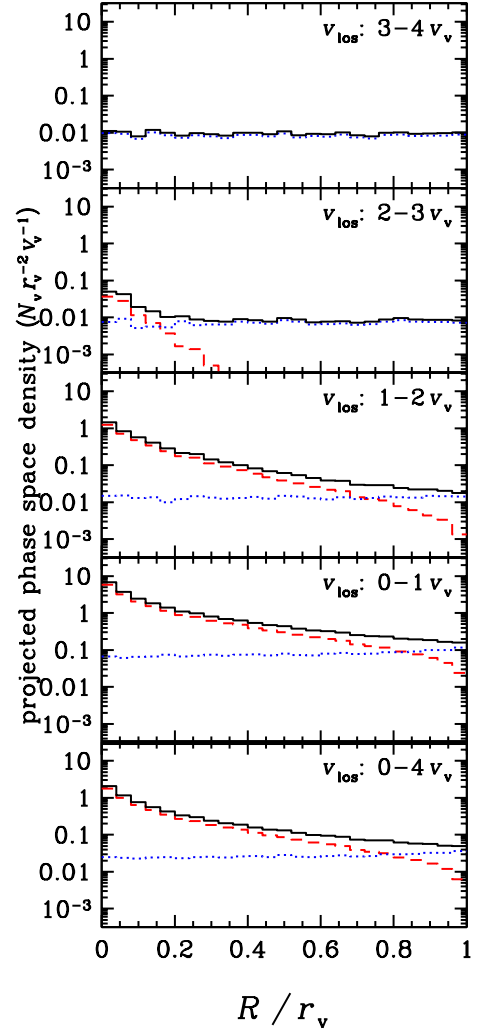}
\caption{Phase space density of stacked virial cone as a function of projected radius in bins of
  absolute line-of-sight velocities
(marked on the upper-right of each plot).
\emph{Dashed (red)} and \emph{dotted (blue) histograms} represent the
halo 
  particles ($r<r_{\rm v}$) and the interlopers ($r>r_{\rm v}$), respectively, while the
  \emph{solid histograms} (artificially moved up by 0.06 dex for clarity)
  represent the full set of particles. 
There are no halo particles at $v > 3\,v_{\rm v}$ (\emph{top plot}).
}
\label{Rhists}
\end{figure}
Figure~\ref{Rhists} shows how the  projected phase space 
density varies with projected radius in different wide velocity bins. 
The halo particles display a negative gradient, i.e. a decreasing surface
number density profile, as expected.
However, one immediately notices that \emph{in all line-of-sight velocity bins, 
the density of interlopers in projected phase space  is roughly
independent of projected radius}. In other words, interlopers have a nearly
flat
surface density profile.

For low velocities, one can notice a small rise of the interloper surface
density at high projected radii. This small rise is a geometric effect: the
line-of-sight distance between the virial sphere and a sphere of $k>1$ virial
radii is 
$k-1$ virial radii at $R=0$ but $\sqrt{k^2-1}$ virial radii at $R=r_{\rm
  v}$, which is $\sqrt{(k+1)/(k-1)}$ times greater.
We will return to this rise in Sect~4.2.

\begin{figure}[ht]
\centering
\includegraphics[width=0.85\hsize,bb=55 10 450 765]{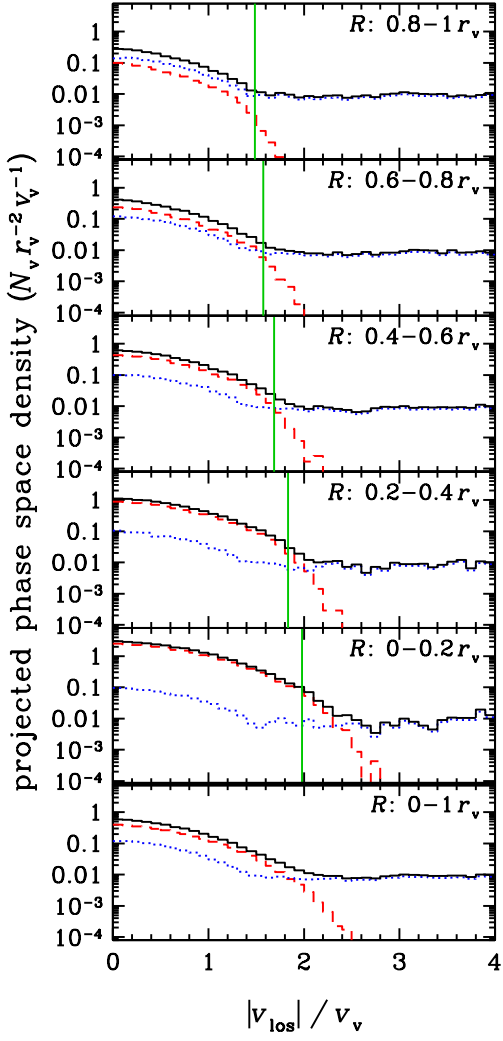}
\caption{Phase space density of stacked virial cone 
as a function of line-of-sight velocity in different
  radial bins (marked on the lower-left of each plot). 
\emph{Dashed (red)} and \emph{dotted (blue) histograms} represent the
halo 
  particles ($r<r_{\rm v}$) and the interlopers ($r>r_{\rm v}$), respectively, while the
  \emph{solid histograms} (artificially moved up by 0.06 dex for
  clarity)
represent the full set of particles.
The \emph{green vertical lines} indicate the $2.7\,\sigma_{\rm los}(R)$ (from
eqs.~[\ref{siglosapx}] and [\ref{siglosconvert}])
  velocity cut for the $c$=4 NFW model with $r_{\rm a}$=$r_{-2}$ ML anisotropy
(eq.~[\ref{betaML}]). 
}
\label{vhists}
\end{figure}
Figure~\ref{vhists} displays the distribution of line-of-sight velocities of
the halo and interloping particles.
The interloper distribution shows a flat component that dominates at large
velocities and a gaussian-like component.
Figure~\ref{vhists} confirms, once more (see Figs.~\ref{Rvzdist} and 
especially \ref{Rhists}) that the
density of interlopers in projected phase space is fairly independent of
  radius.

The total surface density of particles in velocity space shows an inflection
point at about $2\,v_{\rm v}$ (bottom plot of Fig.~\ref{vhists}). Kinematical
modelers attempt to throw out the high-velocity interlopers by identifying
this gap by eye \citep{KG82,LM03} or automatically \citep{Fadda+96}
or by rejecting $3\,\sigma$ outliers, either
using a global criterion \citep{YV77} or a local one (e.g. \citealp{Lokas+06,WL10}).
Interestingly, the $3\,\sigma_v$ criterion was first motivated on statistical
grounds, but the $v \simeq 2.0\,v_{\rm v}$ inflection point one sees in the plots of
Figure~\ref{vhists} happens to correspond to $\simeq 3\,\sigma_v$.
In other words, \emph{the $3\,\sigma_v$ criterion is not only a consequence of
statistics, but also motivated by the 
combination of cluster dynamics and cosmology.}

While visual attempts to separate interlopers from halo particles in projected
phase space look for gaps in the line-of-sight velocity distribution, 
the different panels of Figure~\ref{vhists} indicate that, on average,
one should not expect such gaps in the projected phase space diagram 
of \emph{stacked} clusters, 
as the number of interlopers also decreases with velocity to
reach a plateau at about $2\,v_{\rm v}$.

\subsection{Universality}

\begin{figure}[ht]
\includegraphics[width=\hsize]{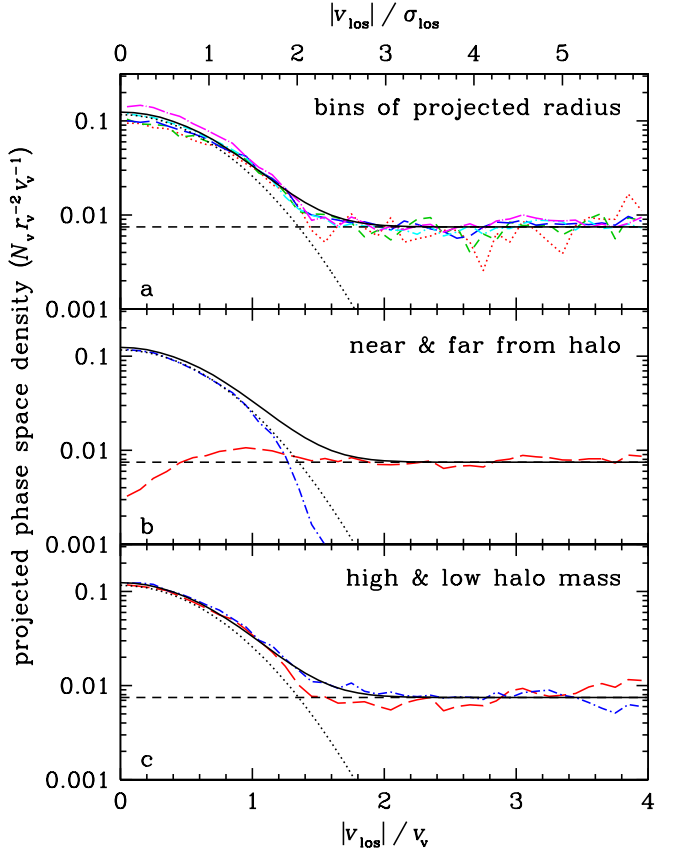}
\caption{Phase space density of interlopers as a function of line-of-sight
  velocity. 
{\bf a} (\emph{top}): Dependence on projected radius: $R/r_{200}$ = 0--0.2, 0.2--0.4,
0.4--0.6, 0.6--0.8, 
and 0.8--1 
(\emph{red dotted}, \emph{green short-dashed}, \emph{blue long-dashed},
\emph{magenta dot-long-dashed} and \emph{cyan dot-long-dashed broken lines}, respectively).
{\bf b} (\emph{middle}): Dependence on 3D radial distance: $1 < r/r_{200} < 8$
(\emph{blue dash-dotted curve}) and $r/r_{200} > 8$ (\emph{red dashed
  curve}).
{\bf c} (\emph{bottom}): Dependence on halo mass: high ($h\,M_{200} > 1.87\times 10^{14}\,M_\odot$,
\emph{red long dashed curve}) and low
($h\,M_{200} < 1.87\times 10^{14}\,M_\odot$, \emph{blue dash-dotted curve}) mass
halos.
Also shown in all three plots 
is the MLE (eqs.~[\ref{gfit}] and [\ref{gfitMLE}])  for the
gaussian (\emph{dotted curve}) and
field 
(\emph{horizontal dashed line}) components,
as well as the
predicted total interlopers (the sum of these two components, \emph{solid
  curve}).
}
\label{vhistint1}
\end{figure}
The three panels of Figure~\ref{vhistint1}  illustrate
the universality of the line-of-sight velocity distribution of interlopers,
in terms of projected radius and halo mass.
Figure~\ref{vhistint1}a confirms that the projected phase space
density of interlopers depends little on projected radius. This will be
quantified later in this sub-section.

As mentioned above, 
to first order, the density of interlopers in projected phase space has
  a constant 
component and a quasi-gaussian component, which we write
\begin{equation}
g\left(R,\left |v_{\rm los}\right|\right) = A\, \exp
\left [-{1\over2}\,\left({\left|v_{\rm los}\right| \over
\sigma_{\rm i}}\right)^2 \right]+ B
\ .
\label{gfit}
\end{equation}
Maximum likelihood estimation (MLE, see appendix~\ref{appmle}) yields
\begin{eqnarray}
\sigma_{\rm i}&=&0.576\pm0.003 \ , \nonumber  \\
A&=&0.1164\pm0.0006 \ , \\
B&=&0.0075\pm0.0001 \ , \nonumber
\label{gfitMLE}
\end{eqnarray}
\noindent where $\sigma_{\rm i}$ is in units of $v_{\rm v}$, while $g$, $A$
and $B$ are in units of 
$N_{\rm v}\,r_{\rm v}^{-2}\,v_{\rm v}^{-1}$, where $N_{\rm v}$ is the number
of particles within the virial sphere.
Figures~\ref{vhistint1}a and \ref{vhistint1}c show that this
gaussian+constant model is an 
excellent fit to the distribution of
interloper line-of-sight velocities.

The origin of these two components is clarified in
Figure~\ref{vhistint1}b: cutting the halo interlopers in two subsamples at
different 3D distances from the halo, one finds that the flat component
corresponds roughly to the interlopers beyond $8\,r_{200}$, while the
quasi-gaussian component corresponds to the closer interlopers ($r_{200} < r
< 8\,r_{200}$).

\begin{figure}[ht]
\centering
\includegraphics[width=\hsize]{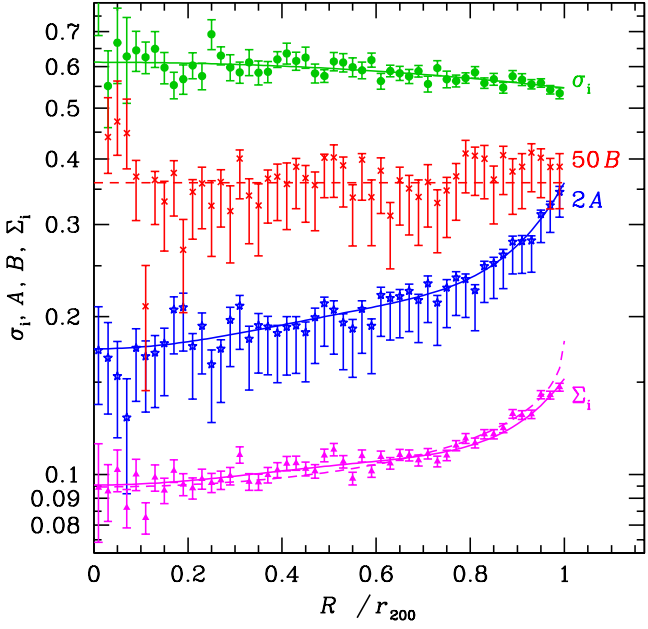}
\caption{Variations of MLE parameters of
equation~(\ref{gfit}) and measured interloper surface density
with projected radius (in units of $v_{\rm v}$ for
$\sigma_{\rm i}$, $N_{\rm
  v}\,r_{v}^{-2}\,v_{\rm v}^{-1}$ for $A$ and $N_{\rm v}\,r_{\rm v}^{-2}$ for
$\Sigma_{\rm i}$).
The \emph{blue} and \emph{green curves} show the fits of
equations~(\ref{AfitvsR}) and (\ref{sigmaifitvsR}), respectively, while the
\emph{red horizontal dashed line} shows the mean of $50\,B$.
The \emph{magenta solid curve} shows the prediction for the interloper
surface density from
equations~(\ref{SigmaivsA}), (\ref{AfitvsR}) and (\ref{sigmaifitvsR}),
respectively, with $B = 0.0075$, while the \emph{magenta dashed curve} shows
the interloper mean surface density predicted (eq.~[\ref{SigmaiNFW}]) from a $c$=4 NFW model.
Small radial bins were chosen to capture the rise of $\Sigma_{\rm i}$ and $A$
near the virial radius.
Errors are from the likelihood ratios in the
MLE fit ($A$, $\sigma_{\rm i}$, and $B$) or Poisson ($\Sigma_{\rm i}$).
}
\label{univvsrad}
\end{figure}

Figure~\ref{univvsrad} shows the radial dependence of the parameters of the
interloper phase space distribution.
We also show the mean surface densities of interlopers measured on the
stacked halo.
The normalization of the gaussian component and the surface density both
increase slowly at small projected radii, but sharply near the virial radius.
A good fit for the normalization and the standard deviation 
of the gaussian component is provided by 
\begin{eqnarray}
A(R) &\!\!\simeq\!\!& {\rm dex} \left [-1.061 + 
0.364\,X^2
- 0.580\,X^4
+ 0.533\,X^6
\right] \ , 
\label{AfitvsR}\\
\sigma_{\rm i}(R) &\!\!\simeq\!\!& 0.612 - 0.0653\,X^2 \ , 
\label{sigmaifitvsR}
\end{eqnarray}
where $X=R/R_{200}$ and ${\rm dex}\,x=10^x$.

The solid magenta curve of Figure~\ref{univvsrad} indicates that
the interloper surface density profile is well recovered from $A$,
$\sigma_{\rm i}$ and $B$ by integrating over the model velocity distribution
(eq.~[\ref{gfit}]) from 0 to $\hat\kappa\,v_{\rm v}$: 
\begin{equation}
\Sigma_{\rm i}(R) = \sqrt{\pi/2}\, A(R) \,\sigma_{\rm i}(R)\,  {\rm
  erf}\left({\hat\kappa\over \sigma_{\rm
  i}(R)\sqrt{2}}\right)+\hat\kappa\, B \ .
\label{SigmaivsA}
\end{equation}
Alternatively,
the surface density profile of interlopers
is also well recovered (dashed magenta curve in Fig.~\ref{univvsrad}) by the
prediction from an NFW model, i.e. as the difference between the standard surface density
integrated to infinity and the surface density limited to the virial sphere:
\begin{equation}
\Sigma_{\rm i}(R) = \Sigma_{\rm NFW}(R) - \Sigma_{\rm NFW}^{\rm sph}(R,R_{200}) \ ,
\label{SigmaiNFW}
\end{equation}
where the expression for $\Sigma_{\rm NFW}^{\rm sph}(R,R_{200})$ is provided
in equation~(\ref{sbrnfwsph}).
According to equation~(\ref{SigmaiNFW}), the rise at radii close to
the virial radius is almost independent of the concentration 
(with relative differences of
less than 4\% at all projected radii). Hence, one can adopt a
unique empirical model for $A$.
These two estimates of $\Sigma_{\rm i}(R)$ are in very good agreement, except that
the difference of the cylindrical and spherical NFW surface density profiles
(eq.~[\ref{SigmaiNFW}])
rises somewhat faster than our fit at projected
radii very close to the virial radius.

The bottom panel of
Figure~\ref{vhistint1} shows that the density of interlopers in projected
phase space  as a function of line-of-sight velocity remains the same for low
and high mass clusters (where we took the dividing line at the median cluster
virial mass of $h\,M_{200} = 1.87\times 10^{14}\,M_\odot$). The interlopers
of high mass
clusters have a cluster-outskirts component whose density in projected phase
space is roughly 15\% lower than that of the low-mass clusters. 

\begin{figure}[ht]
\centering
\includegraphics[width=\hsize]{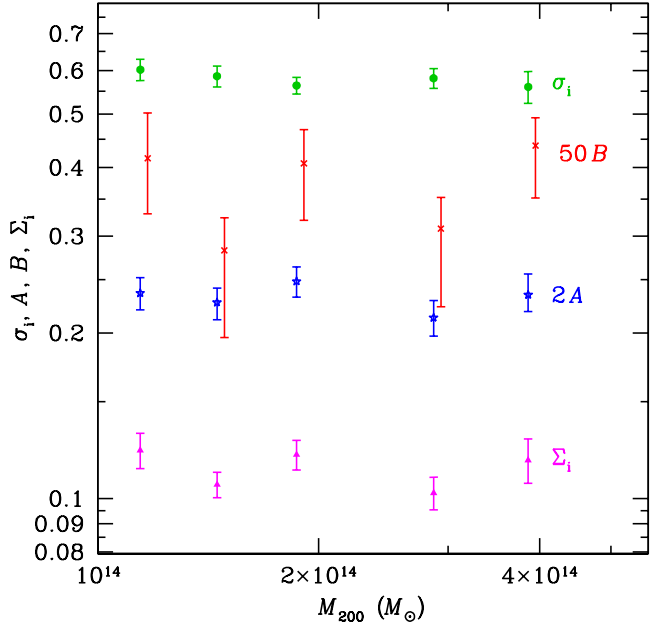}
\caption{Variations of MLE parameters of equation~(\ref{gfit}) 
with halo mass (in units of $v_{\rm v}$ for
$\sigma_{\rm i}$, $N_{\rm v}\,r_{\rm v}^{-2}$ for $\Sigma_{\rm i}$ and $N_{\rm
  v}\,r_{v}^{-2}\,v_{\rm v}^{-1}$ for $A$ and $B$).
Errors are from 100 bootstraps on the 93 halos. The symbols for $50\,B$ are
displaced by +0.01$\,$dex for clarity. }
\label{univvsmass}
\end{figure}
Figure~\ref{univvsmass} provides a closer look at the variation with halo
mass of the parameters of equation~(\ref{gfit}).
The best fitting logarithmic slopes, obtained by
least-squares fits to the points shown in Fig.~\ref{univvsmass} are 
$-0.042$$\pm$$0.030$, $-0.037$$\pm$$0.069$, $0.155$$\pm$$0.225$ and $-0.044$$\pm$$0.104$ for
$\sigma_{\rm i}$, $A$, $B$, and 
$\Sigma_{\rm i}$, respectively.
The 90\% confidence lower limits on the slopes $\d \ln X/\d \ln M$ are thus
$-0.08$, $-0.13$, $-0.17$ and $-0.18$ for $X=\sigma_{\rm i}$, $A$, $B$, and
$\Sigma_{\rm i}$, respectively, while the 90\% upper limits are less than 0.1
except for $B$ (where is it 0.5).
These shallow limits to the logarithmic slopes illustrate the
near-universality of the distribution of interlopers in projected phase
space. 

However,
the universality of the distribution of interlopers in projected phase space
may hide an important level of cosmic variance.
Performing MLE for parameters $\sigma_{\rm i}$, $A$, and $B$,
for each of the 93 clusters, each viewed in turn along each of three
orthogonal viewing axes, we find standard deviations of
\begin{equation}
\sigma(\log \sigma_{\rm i}) = 0.11 \ , \quad \sigma(\log A) = 0.23 \ , \quad \sigma(\log
B) = 0.40 \ .
\label{sigmalogs}
\end{equation}
So, while the dispersion $\sigma_{\rm i}$ of the gaussian component of the interloper
velocity distribution is fairly constant (29\% typical variations) 
from one cluster to the next, there
is more scatter in the normalization $A$ of the gaussian component (factor 1.7
typical variations) and a large scatter in the flat component $B$ (factor 2.5
typical variations).

\section{Interloper removal}
\label{removal}

We remove the interlopers of the stacked cluster proceeding along similar
lines as \cite{Lokas+06} (see also \citealp{Wojtak+07}), by clipping the
velocities beyond $\kappa$ times the \emph{local} line-of-sight velocity dispersion.
We assume that our stacked cluster has an
NFW profile (eq.~[\ref{rhoNFW}]), or, alternatively, an \cite{Einasto65} profile:
\begin{equation}
\nu(r) = {(2m)^{3m}\,\over m\,
  \gamma(3m,2m)}\,\exp\left[-2m\,\left({r\over r_{-2}}\right)^{1/m}\right]
\,\left[{M\left(r_{-2}\right)\over 4\,\pi\,r_{-2}^3}\right] \ ,
\label{rhoE}
\end{equation}
where
$\gamma(a,x)$ is the incomplete gamma
function.
The density model of equation~(\ref{rhoE}) fits
the density profiles of $\Lambda$CDM halos even better than the
NFW model (as first discovered by \citealp{Navarro+04}), at the expense of an
additional parameter, $m$.

The velocity cut requires an estimate of the line-of-sight velocity dispersion profile of the
halo. One could measure this in bins of projected radius, iteratively
rejecting the outliers. This gives a profile that shows important radial 
fluctuations, and we would need to either smooth the profile or fit a smooth
analytical function to it.

Instead, we choose to predict the line-of-sight velocity dispersion profile
given the typical density and velocity anisotropy profiles of halos.
The line-of-sight velocity dispersion profile can be written \citep{ML05b}
\begin{equation}
\sigma_{\rm los}^2 (R) = {2\over \Sigma(R)}\,
\int_R^\infty K\left({r\over R},{r_{\rm a}\over R}\right)\,\nu(r) \,v_c^2(r) \,\d r \ ,
\label{siglossq}
\end{equation}
where $v_c(r) = \sqrt{GM(r)/r}$ is the circular velocity profile, 
$r_{\rm a}$ is the anisotropy radius,
while  
the dimensionless kernel $K$ is
\begin{equation}
K(u) = \sqrt{1-{1\over u^2}} \ ,
\label{Kiso}
\end{equation}
for isotropic orbits \citep{Tremaine+94,PS97} and
\begin{equation}
K(u,u_a) = \left \{
\begin{array}{ll}
\displaystyle
{1/2 \over u_a^2-1}\,\sqrt{1-{1\over u^2}}
+ 
\left ({1+{u_a \over u}}\right) \cosh^{-1} u \\
\displaystyle 
\mbox{} \quad \!-\! \hbox{sgn}\left (u_a\!-\!1\right)
u_a {u_a^2\!-\!1/2\over \left (u_a^2\!-\!1\right )^{3/2}}
\left (1\!+\!{u_a\over u} \right )\\
\displaystyle
\mbox{} \qquad \times \hbox{C}^{-1} \left ({u_a u + 1\over u+u_a} \right )
& \!\!(u_a\!\neq\!1) \,,\\
\\
\displaystyle
\left ({1\!+\!{1 \over u}}\right)\,\cosh^{-1}\!u
-{1\over6}\left ({8\over u}\!+\!7\right)\!\sqrt{u\!-\!1\over u\!+\!1}
& \!\!(u_a\!=\!1) \,,\\
\end{array}
\right.
\label{KML}
\end{equation}
where 
\begin{equation}
\hbox{C}^{-1} (x) = \left \{ 
\begin{array}{ll}
\cos^{-1} x &\hbox{ for } u_a < 1\\
\cosh^{-1} x &\hbox{ for } u_a > 1\\
\end{array}
\right.
\label{KC}
\end{equation}
\citep{ML05b} for the anisotropy profile
\begin{equation}
\beta(r)= {1\over2}\,{r\over r+r_{\rm a}} \ ,
\label{betaML}
\end{equation}
which \cite{ML05b} (hereafter, ML) found to be a good fit to the
anisotropy profiles of $\Lambda$CDM halos.

Since the space density model enters equation~(\ref{siglossq}) expressing
$\sigma_{\rm los}(R)$ (through the tracer density $\nu$ and the total mass
$M$, which are related since we are considering single component mass
models),
we first need to determine the best fitting model to the distribution of
particle radii in the stacked halo:
we performed MLE of the NFW and Einasto models to the
distribution of 3D radii of our stacked cluster. The minimum radius was chosen
as 
$0.03\,r_{200}$ to avoid smaller radii, since our halo centers appear to be
uncertain to about 1\% of the virial radius.
We varied the outer radii of the fit, starting at $r_{200}$. 
Since we will later fit the surface density profile out to the virial radius
and beyond, we need to
remember that the space radii extend beyond the maximum projected radius of
the future surface density fits. 
So we also
performed 3D fits beyond the virial radius: at $1.35\,r_{200}$ (which
corresponds to the radius where $\Delta = 100$, i.e. the largest radius where
the halos should be close to
virial equilibrium), and $3\,r_{200}$ for a broader view of halos far beyond $r_{200}$.

When 
\cite{Prada+06} fit the density profiles of $\Lambda$CDM halos out to 
$2.7\,r_{200}$, they found them to be well
approximated by the sum of an Einasto model and a constant
term $\rho_{\rm bg} = \Omega_{\rm m}\,\rho_{\rm c}$.
We therefore also experimented with the addition of 
a constant background component of density equal to the
density of the Universe. In virial units, this background is
expected to be equal to $\hat \nu_{\rm bg}=3\,\Omega_{\rm m}/(4\pi \Delta) = 3.6\times 10^{-4}$ (with
$\Omega_{\rm m}=0.3$ and $\Delta=200$). 

\begin{table}[ht]
\caption{MLE fits to the radial distribution of the stacked halo
\label{c3d}}
\centering
\begin{tabular}{lllcl}
\hline\hline
Model & $r_{\rm max}$ & \multicolumn{1}{c}{$100\,\hat\nu_{\rm bg}$} & $c$ & \multicolumn{1}{c}{$P_{\rm KS}$} \\
\hline
NFW & 1 & 0 & 4.08$\pm$0.05$\pm$0.17 & 0.0021\\ 
NFW & 1 & (0.036) & 4.10$\pm$0.05$\pm$0.17 & 0.0023\\ 
NFW & 1 & 0.416 & 4.26$\pm$0.07$\pm$0.23 & 0.0069\\ 
Einasto & 1 & 0 & {\bf 4.00$\pm$0.05$\pm$0.17} & 0.31\\ 
Einasto & 1 & (0.036) & 4.02$\pm$0.05$\pm$0.17 &  0.3\\ 
Einasto & 1 & 0.002 & 4.00$\pm$0.07$\pm$0.20 & 0.31\\ 
NFW & 1.35 & 0 & 4.20$\pm$0.05$\pm$0.14 & 0.0023\\ 
NFW & 1.35 & (0.036) & 4.23$\pm$0.05$\pm$0.14 & 0.0021\\ 
NFW & 1.35 & 0.002 & 4.20$\pm$0.04$\pm$0.15 & 0.002\\ 
Einasto & 1.35 & 0 & {\bf 4.14$\pm$0.04$\pm$0.13} & 0.016\\ 
Einasto & 1.35 & (0.036) & 4.17$\pm$0.04$\pm$0.13 & 0.0058\\ 
Einasto & 1.35 & 0.002 & 4.14$\pm$0.04$\pm$0.14 & 0.015\\ 
NFW & 3 & 0 & 4.80$\pm$0.04$\pm$0.24 &    0\\ 
NFW & 3 & (0.036) & 5.01$\pm$0.04$\pm$0.24 &    0\\ 
NFW & 3 & 0.001 & 4.80$\pm$0.03$\pm$0.19 &    0\\ 
Einasto & 3 & 0 & 4.26$\pm$0.03$\pm$0.22 & $<$$10^{-15}$\\
Einasto & 3 & (0.036) & {\bf 4.50$\pm$0.03$\pm$0.23} &  $<$$10^{-10}$\\
Einasto & 3 & 0.042 & 4.54$\pm$0.04$\pm$0.13 & $<$$10^{-11}$\\ 
\hline
\end{tabular} 
\tablefoot{The Einasto models are for index $m=5$. Column (2) is the maximum
radius for the fits (the minimum radius 
is set to $0.03\,r_{200}$). Columns (3) and (4) are respectively
the background
density (multiplied by 100, fixed if in parentheses)
and concentration, while column (5) is the Kolmogorov-Smirnov
test probability that the model is consistent with the distribution of radii.
Virial units (with $\Delta=200$) are used for $r_{\rm max}$ and $\hat\nu_{\rm
  bg}$ ($r_{200}$ and 
$N_{200}/r_{200}^2$, respectively), while $c=r_{200}/r_{-2}$, where
$r_{-2}$ is the radius of density slope equal to $-2$. The outer radius of
$1.35\,r_{200}$ 
corresponds to the virial radius, $r_{100}$.
Errors are $1\,\sigma$: the first are the statistical errors 
from likelihood ratios and the second are from cosmic variance
using 100 cluster bootstraps.
Concentrations in bold face are those providing the highest $P_{\rm KS}$
tests for given $r_{\rm max}$.}
\end{table}

Table~\ref{c3d} shows 
the resulting best-fit concentrations\footnote{In this paper, concentrations
  refer to $r_{200}/r_{-2}$.} obtained by MLE fits of NFW and $m$=5
Einasto models,
plus an optional fixed or free constant background, to
the distribution of 3D 
radii of the stacked halo.\footnote{We also experimented with free index Einasto models: we
  generally found that the best fit index was in the range $4.6 < m < 5.2$.}
The best-fit concentrations increase with the maximum allowed projected radius
when the NFW model is
used. This indicates the inadequacy of the NFW model at large radii, as it
fails to capture the steepening of the slope of the density profile  beyond
the virial radius
\citep{Navarro+04}. In fact, a Kolmogorov-Smirnov test (last column in
Table~\ref{c3d})
indicates that the
NFW model is not an adequate representation of the distribution of radii,
whether a constant background is added or not, regardless of the maximum
radius used in the fit.

On the other hand, the concentration of the Einasto model appears to be
somewhat less dependent of the outer radius, as also noted by \cite{Gao+08}.
At $0.03<r/r_{200}<1$, the Einasto model is an adequate representation of the
distribution of radii. At $0.03<r<r_{100}=1.35$, 
the $m$=5 Einasto model is inconsistent with the distribution of radii, but
not by a large amount.
However, the distribution of radii extending far beyond the virial radius,
$0.03<r/r_{200}<3$ is not consistent with either NFW or $m$=5 Einasto 
models.\footnote{For
$r_{\rm max} =3\,r_{200}$, the KS test showed that the free $m$ Einasto model
with free or fixed background  (or without any) failed to provide an adequate
representation of the distribution of radii.}

The inclusion of the background in the fits leads to even higher
concentrations when the maximum radius considered is $3\,r_{200}$.
\begin{figure}[ht]
\includegraphics[width=\hsize]{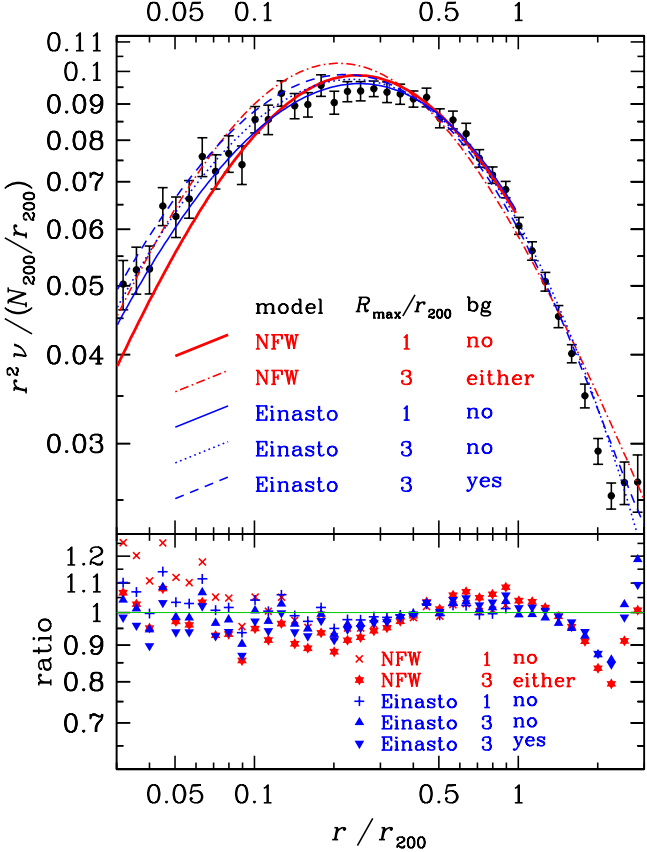}
\caption{\emph{Top}: Space density profile (multiplied by $r^2$) 
of the stack of the 93 halos, with best maximum
  likelihood fits for $r/r_{200}$ from 0.03 to 1 (\emph{solid}), and 3
  (\emph{dotted} without background, \emph{dashed curves} with best-fit
  background, see Table~\ref{c3d}) for the
  NFW (\emph{red}) and Einasto (\emph{blue}) models. The best fit NFW models to
  $3\,r_{200}$ with and without background are
  indistinguishable (see Table~\ref{c3d}). The errors are from 100 cluster
  bootstraps.
\emph{Bottom}: Ratios of measured to fit densities.
}
\label{rhoofr}
\end{figure}

The density profile of the 93 stacked clusters is shown in
Figure~\ref{rhoofr} for maximum fit radii of $r_{\rm max} = 1$ and
$3\,r_{200}$.
The maximum likelihood NFW model produces clearly worse fits to the density
profile than the maximum likelihood Einasto model for $r < 
3\,r_{200}$, while the NFW model reproduces better the measured density
profile at $r = 3\,r_{200}$. As clearly seen in the bottom panel of
Figure~\ref{rhoofr}, neither model is adequate near
$2\,r_{200}$, even when considering the cosmic variance measured by our cluster
bootstraps (see the error bars in the top panel of Fig,~\ref{rhoofr}).

In most of what follows, we restrict our analysis to $R < r_{200}$. For these
analyses, we adopt
the $c=4$, NFW and $m$=5 Einasto models, as these models are simple and the
latter is consistent
with the radial distribution without and with a constant background.

\begin{figure}[ht]
%
\includegraphics[width=\hsize]{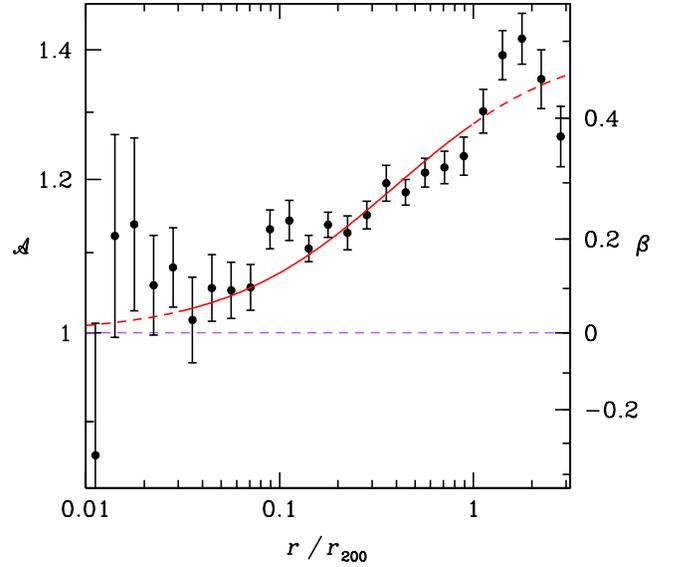}
\caption{Velocity anisotropy profile (including streaming motions:
  eqs.~[\ref{betadef}] and [\ref{betapdef}]) of the stack of the 93 halos. 
The error bars are from 100 bootstraps on the 93 halos.
The \emph{curve} shows the weighted $\chi^2$ fit
  (in the range $0.03 < r/r_{200} < 1$) of the
  Mamon-{\L}okas anisotropy (2005b) model (eq.~[\ref{betaML}]) with $r_{\rm a} =
  0.27\,r_{200}$ (the \emph{solid portion} of the curve highlights the
  region where the fit was performed). 
The \emph{purple dashed horizontal line} indicates the fully isotropic case.
\label{betafig}} 
\end{figure}
The radial profile of  velocity anisotropy (eq.~[\ref{betadef}]) 
of the stacked cluster, shown in Figure~\ref{betafig},
is reasonably well fit by (reduced $\chi^2=1.4$) the ML
model\footnote{Other anisotropy models such as constant and Osipkov-Merritt
  \citep{Osipkov79,Merritt85_df} 
  produce much worse best fits (reduced $\chi^2 \simeq 5$ and 20, respectively).}
(eq.~[\ref{betaML}]), 
with anisotropy radius (where $\beta$ reaches its half-value of 1/4) $r_{\rm a} =
0.275\pm0.020\,r_{200}$, as found with a weighted $\chi^2$ fit for $\log
r/r_{200}$ between $-1.5$ and 0. With a concentration parameter $c =
4.0$, the anisotropy radius is $r_{\rm a} \simeq 1.1\, r_{-2}$, i.e. \emph{the velocity
  anisotropy 
reaches its value intermediate between the center and the outer regions close to
the radius of density slope $-2$}.

We now measure the velocity dispersion of the stacked profile using different
schemes for interloper removal to find a scheme that produces a velocity
dispersion profile (on the data with Hubble flow and the velocity cut)
in agreement with the predictions (with no Hubble flow nor velocity cut) 
for the $c=4.0$ NFW and Einasto
models whose density profiles we just fit in 3D.
Since the surface density profile enters equation~(\ref{siglossq}), and no
analytical formula is known for the Einasto model, we derived an accurate
approximation for the Einasto surface density profile (for a large range of
projected radii and of indices $m$) in
Appendix~\ref{appsdensEinasto} (eqs.~[\ref{Sigmaapx}] and [\ref{Rapx}]).

\begin{figure}[ht]
\includegraphics[width=0.95\hsize]{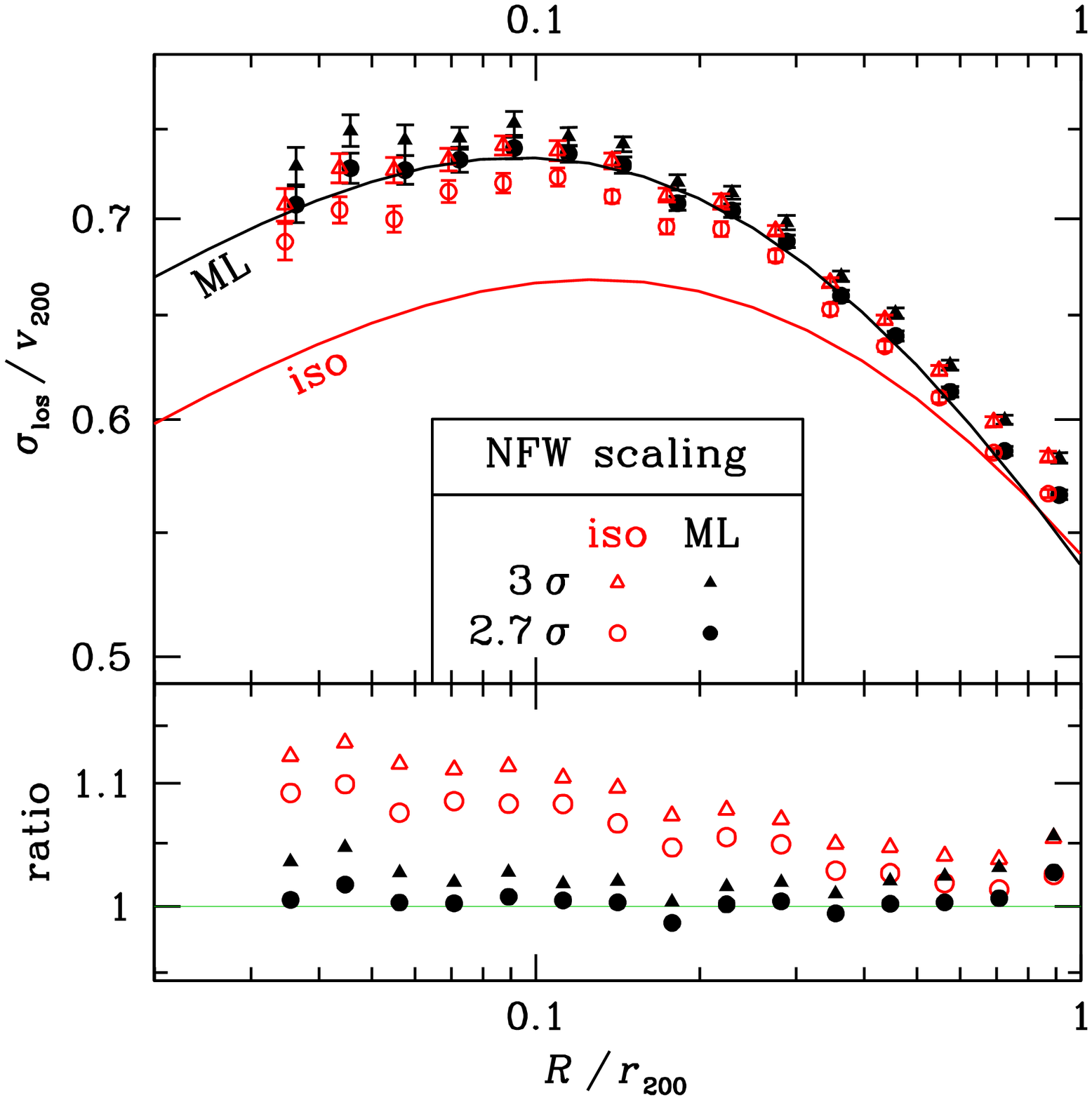}
\includegraphics[width=0.95\hsize]{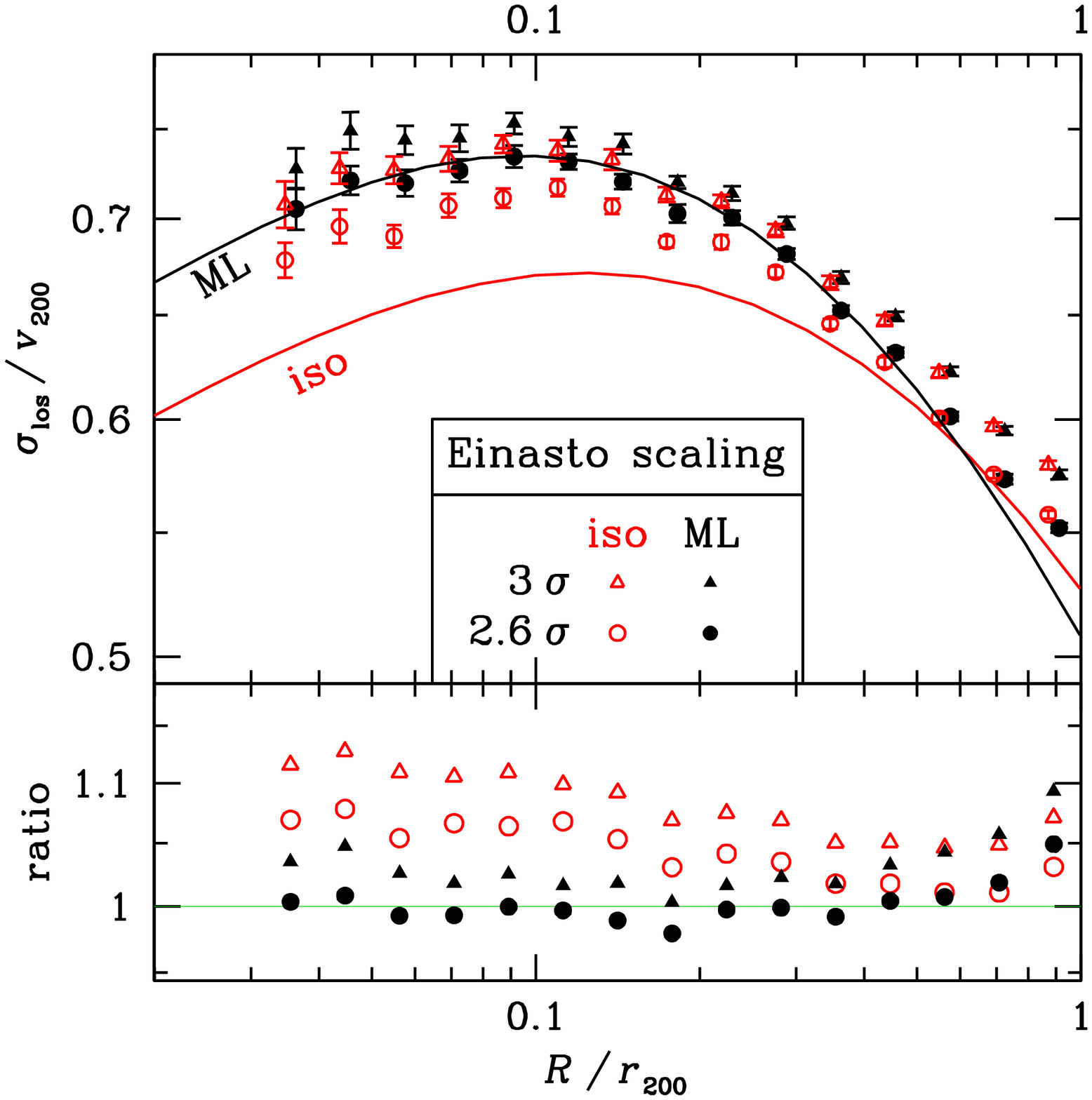}
\caption{Line-of-sight velocity dispersion profiles of the stacked virial
  cone,  cutting at 3 (\emph{triangles}) or 2.7 (\emph{circles})
$\sigma_{\rm los}(R)$, assuming the best fitting ($c=4.0$) NFW
(\emph{top}) or $m\!=\!5$ Einasto (\emph{bottom}) model
with isotropic 
velocities (\emph{red open symbols}) or slightly radial ML
anisotropy (eq.~[\ref{betaML}])  with $r_{\rm a}
= r_{-2}$ (\emph{black filled symbols}).
The error bars are from 100 bootstraps on the particles within each bin of
projected radii. 
For clarity, the isotropic and ML symbols are shifted by 0.01 dex leftwards and
rightwards, respectively.
The \emph{curves} show the predicted line-of-sight velocity dispersions (for
no Hubble flow,
eq.~[\ref{siglossq}]) assuming
isotropy (\emph{red curve}, eq.~[\ref{Kiso}]) or  
ML  
anisotropy  with $r_{\rm a} = r_{-2}$
(\emph{black curve}, using eq.~[\ref{KML}]).
The ratios of measured to predicted $\sigma_{\rm los}(R)$ are shown in the
lower frames of each plot
(same colors and symbols as upper frames).
\label{siglos3}}
\end{figure}

The red open triangles in the top panel of 
Figure~\ref{siglos3} show the line-of-sight
velocity dispersion profiles ``measured'' with our standard velocity cut at 3
times the predicted 
isotropic line-of-sight velocity dispersion (hereafter $\sigma_{\rm
  los}^{\rm iso}$) for an NFW model with concentration $c=4.0$ (as
measured in 3D, see above).
In comparison, 
$\sigma_{\rm los}^{\rm iso}$ 
(red solid curve in the top panel of 
Fig.~\ref{siglos3}) is typically 10\% lower than the ``measured'' velocity
dispersion profile for radii $R < 0.1\,r_{200}$.
This discrepancy is decreased to 4\% if one compares the measured velocity
dispersions after clipping at 3 times the line-of-sight velocity dispersion, 
computed with the ML anisotropy (hereafter $\sigma_{\rm los}^{\rm ML}$, 
eqs. ~[\ref{siglossq}] and [\ref{KML}],
black filled triangles)
to $\sigma_{\rm los}^{\rm ML}$ 
(black solid
curve in the top panel of Fig.~\ref{siglos3}). 
This suggests that the $3\,\sigma$ clipping generally used is too liberal. 
A near perfect match (typically better than 1\% for $R < 0.8\,r_{200}$) is
obtained by 
cutting at 
$2.7\,\sigma_{\rm los}^{\rm ML}$ (black  filled 
circles vs. black solid curve in the top panel of Fig.~\ref{siglos3}).

When the $m$=5 Einasto model is used to compute $\sigma_{\rm los}(R)$ before
applying the velocity cut, the best
match between the measured and predicted line-of-sight velocity dispersion
profiles is for a cut at $\kappa = 2.6$ (bottom panel of Fig.~\ref{siglos3}).
 
The
Hubble flow (HF) causes a shallower slope at projected radii close to the virial
radius (one notices in both panels of Fig.~\ref{siglos3} an inflection point
in the measured profiles [filled 
  circles] of
$\log \sigma_{\rm los}$ vs. $\log R$ near half a virial radius).
Indeed, we obtained results similar to those of Figure~\ref{siglos3} when
we did not
incorporate the HF to the peculiar velocities of the simulation:
the measured $\sigma_{\rm los}(R)$ fell more sharply, with no inflection point, even
somewhat more sharply than predicted by the
Einasto model (because the velocity anisotropy without the HF is more radial at $\simeq
4\,r_{200}$ in comparison with the case where the HF is
incorporated, where $4\,r_{200}$ roughly corresponds to the turnaround radius
where the velocities are mostly tangential).
So, although the steeper Einasto density profile ought to catch better the
steeper line-of-sight velocity dispersion profile at large projected radii, 
the NFW
model performs slightly better, because its shallower 
line-of-sight profile mimics better the effects of the Hubble flow.

In summary, Figure~\ref{siglos3} indicates that if one wishes to recover the
correct line-of-sight velocity dispersion profile, 
\emph{one should use 2.6 or 
$2.7\,\sigma$ clipping instead of $3\,\sigma$ clipping,
  where the line-of-sight velocity dispersion is either measured or modeled
  with anisotropic velocities}.

The choice of model and $\kappa$ is not obvious. We prefer the NFW model, as
it is simpler and, with $\kappa=2.7$, it presents a slightly better match between measured and
predicted line-of-sight velocity dispersion profiles than does the $m$=5
Einasto model (compare the ratios 
of measured to predicted $\sigma_{\rm los}(R)$ in both plots of
Fig.~\ref{siglos3}, especially at large radii). 

Mass modelers of clusters may wish to avoid performing the integral of
equation~(\ref{siglossq}) with the kernel of equations~(\ref{KML}) and
(\ref{KC}).  
The line-of-sight velocity dispersion profile (eq.~[\ref{siglossq}]) for the
NFW and $m$=5 Einasto models with  
ML anisotropy with $r_{\rm a}
= r_{-2}$ can be approximated as
\begin{eqnarray}
&&{\sigma_{\rm los}(R)\over \sqrt{GM(r_{-2})/r_{-2}}}
\simeq
{\rm dex} 
\left \{
\sum_{i=0}^7 a_i \left[\log_{10} \left({R \over r_{-2}}\right)\right]^i
\right\}
\label{siglosapx}
\end{eqnarray} 
where the coefficients are given in Table~\ref{sigapxcoeffs} for both models.
These two approximations are accurate to better than 0.5\% (rms) 
for $0.0032 < R/r_{-2} < 32$.
\begin{table}[ht]
\caption{Coefficients for $\sigma_{\rm los}(R)$  approximation
  (eq.~[\ref{siglosapx}]) 
\label{sigapxcoeffs}}
\centering
\begin{tabular}{lll}
\hline\hline
& \multicolumn{1}{c}{NFW} & $m$=5 Einasto \\
\hline
$a_0$ & --0.1478 & --0.1520 \\
$a_1$ & --0.1109 & --0.1242 \\
$a_2$ & --0.1357 & --0.1637 \\
$a_3$ & \ \ 0.001948 & --0.01688 \\
$a_4$ & \ \ 0.02317 & \ \ 0.01892 \\
$a_5$ & \ \ 0.0006310 & \ \ 0.001844 \\
$a_6$ & --0.003234 & --0.002044 \\
$a_7$ & --0.0006370 & --0.0004103 \\
\hline
\end{tabular} 
\end{table}
Then, one can write
$\sigma_{\rm los}(R)$ in terms of $v_{\rm v}$ using
equation~(\ref{siglosapx}) and 
\begin{equation}
{v_{\rm v}^2\over GM(r_{-2})/r_{-2}} = 
\left \{
\begin{array}{ll}
\displaystyle
{\ln (c+1)-c/(c+1)\over (\ln
    2-1/2)\,c} & \hbox{(NFW)} \ , \\
\\
\displaystyle
{\gamma\left(3m,2m\, c^{1/m}\right) \over \gamma(3m,2m)\,c}
& \hbox{(Einasto)}
\ ,
\end{array}
\right.
\label{siglosconvert}
\end{equation}
(e.g., \citealp{NFW96} for NFW and trivially derived from the Einasto mass
profile first derived by \citealp{ML05a}).

In the absence of information on the mass profile (e.g. from X ray
observations), neither the concentration parameter, $c$, nor the scale radius
$r_{-2}$ are known, so one has to work iteratively, first guessing a plausible
value of $c$, applying the velocity filter, then re-estimating  $c$ from the
data and re-applying the velocity filter.
This process should converge in a one or two iterations.

\section{Interloper statistics after the velocity cut}
\label{ilopstatsaftervcut}
We now show the statistics of interlopers after our adopted velocity cut.
The motivation is to allow observers to compare with their own data. With the
velocity cut ($\kappa = 2.7$), 
the line-of-sight distances are now effectively limited to $\pm
17\,r_{200}$ from the center of the stacked halo (eq.~[\ref{rmax2}]). 
The
qualitative features of the projected phase space distribution are robust to
variations of the method to cut the velocities.

\label{statswohivilop}
Figure~\ref{phasespacewcuts} shows the 
velocity cut at $\pm$$2.7\,\sigma_{\rm los}(R)$ --- with our adopted NFW model
with ML anisotropy with anisotropy radius $r_{\rm a} =
r_{-2}$ --- on top of the projected phase 
space (using eqs.~[\ref{siglosapx}] and [\ref{siglosconvert}]). 
Only 0.4\% of the halo particles are rejected by the $2.7\,\sigma$
velocity cut,
which is a low enough fraction that the shot noise in the 
structural and kinematical modeling
is not significantly increased.
The velocity cut in Figure~\ref{phasespacewcuts} seems very
reasonable as it is close to optimizing the completeness of the selection of particles
within the virial sphere.
However, less than 17\% of the interlopers are identified as such by the
velocity cut. Therefore, 
\emph{the great majority of interlopers cannot be removed by a velocity cut}. 

The green curves in Fig.~\ref{Rvzdist} show the velocity cut on top of the phase space density
distribution.
The velocity cut appears to occur in a region where the interloper phase
space density is roughly constant.

This can be seen in a clearer fashion in Figure~\ref{vhists}, where the
velocity cut is shown as green vertical lines.
While the highest velocity interlopers are removed by the velocity cut,
there remains signs of the field component, which we
had identified with particles beyond 8 virial radii, at low ($R <
0.4\,r_{200}$) projected radii.

\begin{figure}[ht]
\includegraphics[width=\hsize]{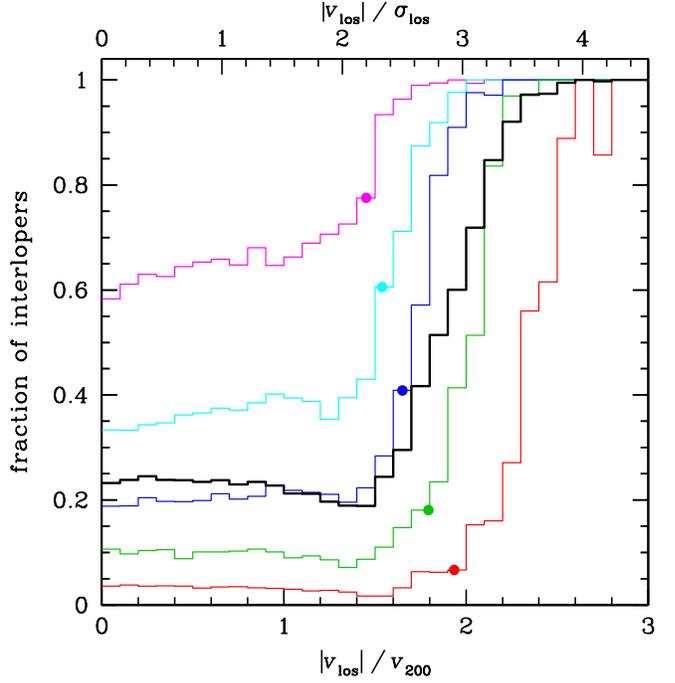}
\caption{Fraction of interlopers as a function of line-of-sight velocity for
  all projected radii $R < r_{200}$ (\emph{thick black histogram}) and in
bins of projected radius: $R/r_{200}$ = 0--0.2, 0.2--0.4, 0.4--0.6, 0.6--0.8,
and 0.8--1 (\emph{thin histograms}), increasing upwards.
The
\emph{filled circles} show 
the $2.7\,\sigma_{\rm los}(R)$ (from
eqs.~[\ref{siglosapx}] and [\ref{siglosconvert}])
  velocity cut for the $c$=4 NFW model with $r_{\rm a}$=$r_{-2}$ ML anisotropy
(eq.~[\ref{betaML}]).
}
\label{ilopfracs}
\end{figure}
The fraction of particles outside the virial sphere is displayed in
Figure~\ref{ilopfracs}. 
Interestingly, 
at $R > 0.8\,r_{200}$ (magenta histogram) the interlopers account for over 60\% of
all particles, regardless of the particle velocity up to the velocity cut
(filled circles).
But even at smaller radii, $0.4 < R/r_{200} < 0.6$, interlopers account for
over 20\% of all particles again for all velocities up to the cut.
So, unless one limits
one's kinematical analysis to very small cluster apertures, one cannot avoid
being significantly contaminated by interlopers. 

While there is no gap in the velocity distribution of particles
(Fig.~\ref{vhists}), 
Figure~\ref{ilopfracs} shows local minima of the interloper fraction for
all bins of projected radii, except the outermost one. Regardless of the
application of a velocity cut, these local minima
occur at lower velocity (1.3 to $1.5\,v_{\rm v}$) than the inflection points of the
interloper density in projected phase space (1.6 to $2.6\,v_{\rm v}$ as seen in
Fig.~\ref{vhists}). 
The local minima occur at velocities that decrease with projected radius,
suggesting that our local $\kappa \,\sigma_{\rm los}$ cut is preferable to a
global one, since $\sigma_{\rm los}(R)$ decreases with $R$ for $R >
0.1\,r_{200}$ (see Fig.~\ref{siglos3}).
These local minima arise because the interloper system has a lower velocity
dispersion than the halo system: $\sigma_{\rm i}=0.58$ (eq.~[\ref{gfitMLE}])
while after the velocity cut the aperture velocity dispersion of the global
stacked virial cone (thus including both halo particles and interlopers)
is $\eta=0.65$,
i.e. 5\% higher than predicted by \cite{MM07} for an isotropic NFW model,
which is not surprising given the radial anisotropy of the halos
(Fig.~\ref{betafig}).

\begin{figure}[ht]
\centering
%
\includegraphics[width=\hsize]{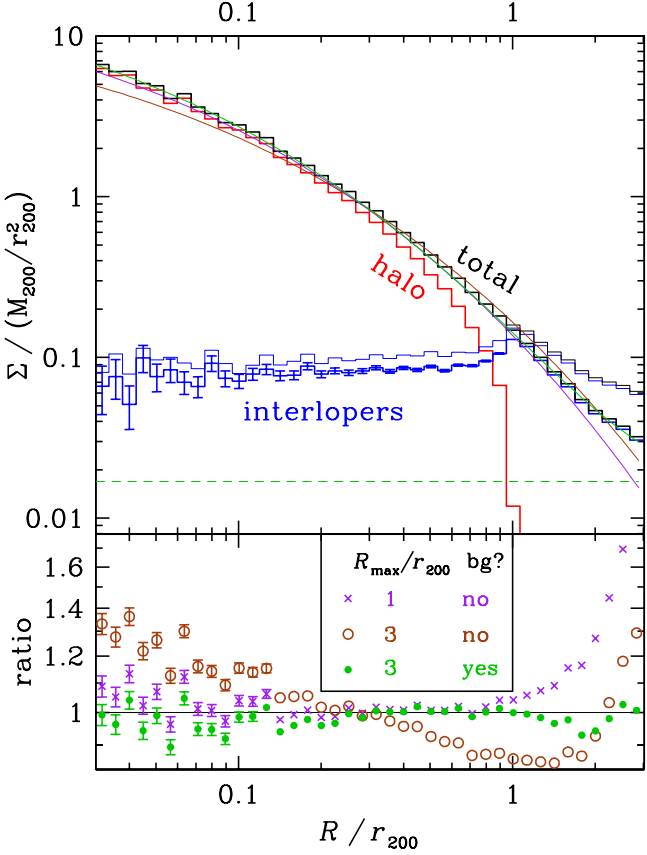}
\caption{\emph{Top panel}: Surface density profile of global stacked cone
(\emph{black histogram}, raised up by 0.02 dex for clarity), 
as well as halo members ($r\leq r_{200}$, \emph{red histogram})
  and interlopers ($r>r_{200}$) before (\emph{thin}) and after
  (\emph{thick blue histograms}) the velocity cut.
Poisson errors are only shown for the interlopers after the velocity cut 
for $R < r_{200}$.
Also shown (\emph{curves}) are maximum likelihood $m$=5 Einasto model fits (after
the velocity cut), in the
range 
$0.03\,r_{200}$ to 1 (\emph{purple}) or 3 (\emph{brown and green})
$r_{200}$, without (\emph{purple and brown}) or with (\emph{green}) an
additional free constant background component (\emph{dashed green line}).
\emph{Bottom panel}: Ratios of measured to fit surface densities (after the
velocity cut). For
clarity, Poisson
errors are only shown if larger than the symbol size.
\label{sdens}}
\end{figure}

The surface density profile of the stacked halo is shown in
Figure~\ref{sdens}.
The surface density profile of the interlopers is flat with small fluctuations
around the mean values 
$\Sigma_{\rm i}=0.114$ and $0.096\,N_{\rm v}\,r_{\rm v}^{-2}$, measured in the
stacked virial cone, respectively before and after
the velocity cut.\footnote{Figure~\ref{sdens} shows interloper surface
  densities that are lower, at $R < 0.7\,r_{200}$, than 0.114 and
  $0.096\,N_{\rm v}\,r_{\rm v}^{-2}$, respectively before and 
  after the velocity cut, but most of the
particles lie within the highest bins of log projected radius.} 
In comparison, our model of the surface density of interlopers
(eq.~[\ref{SigmaivsA}]) combined with our MLE values for $A$, $B$ and
$\sigma_{\rm i}$ yields mean interloper densities of 0.114 and $0.094\,N_{\rm
  v}\,r_{\rm v}^{-2}$, 
respectively before and after the $\kappa=2.7$ ($\hat\kappa = 1.76$ with
$\eta=0.65$) velocity cut. The general agreement is excellent.

Note that, at projected radii beyond the virial
radius, all particles are interlopers,
so the surface density of interlopers is not constant but decreases, to first
order, 
as the NFW or Einasto models.
While the total surface density profiles of the popular NFW and Einasto
models for $\Lambda$CDM halos are convex in log
surface density vs. log projected radius, an important additional background
term in the surface density would lead to an inflection point and subsequent
concavity at some radius.
Such a feature would lead to poor fits of single NFW or Einasto profiles.
Now, within the virial radius, no such inflection point and outer
concavity are seen in 
 Fig.~\ref{sdens} for the \emph{total}
surface density
profile.  
However, extending the surface density profile out to three virial radii, as
illustrated in Figure~\ref{sdens}, one
does see the inflection point of the total surface density profile (near
$1.6\,r_{200}$ after the velocity cut and right at $r_{200}$ before). 
This points to an additional background of surface density.
This requirement for the additional background component 
is confirmed by the fairly flat
ratios of data over model (bottom panel of Fig.~\ref{sdens}) for the case
where the background is fit, in comparison 
with larger residuals for $R > r_{200}$ for the
fits without a background.

Such a background  is expected, since the density
profiles of $\Lambda$CDM halos is the sum of an Einasto model and a constant
background corresponding to the mean density of the Universe
\citep{Prada+06}.
Integrating along the line-of-sight (eq.~[\ref{projec}]) within the sphere of
radius $r_{\rm max}$ (eq.~[\ref{rmax2}]), one
deduces that the surface density profile of (foreground/)background structures is
\begin{equation}
\Sigma_{\rm bg}(R) =  2\,\Omega_{\rm m}\,\rho_{\rm
  c}\,\sqrt{r_{\rm max}^2-R^2}
\ .
\label{SigmaPrada}
\end{equation}
Since $r_{\rm max} \gg r_{\rm v}$ (eq.~[\ref{rmax2}]),
$\Sigma_{\rm bg}$
is roughly constant for $R \la r_{\rm v}$:
\begin{equation} 
\Sigma_{\rm bg} \simeq 2\,\Omega_{\rm m}\,\rho_{\rm
  c}\,r_{\rm max} \ ,
\label{Sigmabg}
\end{equation}
which in dimensionless virial units ($N_{\rm v}\,r_{\rm
  v}^{-2}$)  
becomes
\begin{equation}
\hat\Sigma_{\rm bg} = {\Sigma_{\rm bg}\over N_{\rm v}/r_{\rm v}^2} 
\simeq {3\over \pi}\,{\eta\,\kappa\,\Omega_{\rm m}\over \sqrt{8\,\Delta}} \ ,
\label{Sigmabgvir}
\end{equation}
using equation~(\ref{rmax}).
With the velocity cut, $\kappa=2.7$ and equation~(\ref{Sigmabgvir}) 
yields $\hat \Sigma_{\rm bg} \simeq 0.0126$ for $\Omega_{\rm m}=0.3$,
$\Delta=200$, and
$\eta=0.65$ (see above).
Without the velocity cut $\hat\kappa = \eta\,\kappa = 4$ and one obtains
$\hat \Sigma_{\rm bg} \simeq 0.0286$.
According to equation~(\ref{SigmaPrada}), the relative drops of
$\Sigma_{\rm bg}$ from $R=0$ to $R_{\rm max}=1$ or $3\,r_{\rm v}$ are 0.2\%
and 1.6\%, respectively, so the approximation of a constant surface density 
background is adequate.

This background corresponds to the velocity-independent component of the
interloper surface phase space density, which
in our model (eq.~[\ref{gfit}]) is the $B$ term, which produces a mean
surface density of
$\hat \Sigma_{\rm bg} = \eta\times 2.7\,B = 0.0132$ (with $B=0.0075$ from
eq.~[\ref{gfitMLE}] and again $\eta=0.65$). This agrees with the 
previous value to within 4\%. This means that the constant field term in the
velocity distribution corresponds precisely to the additional halos outside
the test halo.
In any event,
\emph{the total surface density of a cosmological structure is the sum
of the surface density of that structure and a constant background}.

\begin{figure}[ht]
\centering
\includegraphics[width=\hsize]{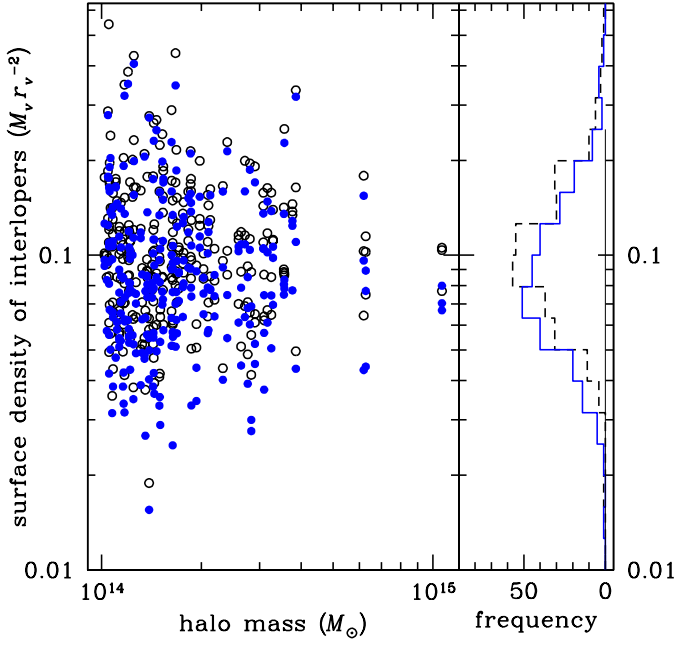}
\caption{Mean interloper surface density (in virial units)
versus halo mass (for $\Delta=200$). 
The \emph{open black} and \emph{filled blue circles} show
the 93
halos each measured along 3 orthogonal viewing directions, before and after 
  the velocity cut, respectively.
The \emph{right panel} provides the frequency of the mean interloper surface
density (with logarithmic bins), before (\emph{dashed black}) and after
(\emph{solid blue histograms}) the velocity cut.}
\label{sdensvsm}
\end{figure}

Although the mean surface density of interlopers is roughly independent of
halo mass (Fig.~\ref{univvsmass}),
it may vary from cluster to cluster.
Figure~\ref{sdensvsm} shows the surface density of each of the 93 halos.
As seen  in Table~\ref{ilopstats},
\begin{table}[ht]
\caption{Statistics of interloper surface densities $\Sigma_{\rm i}$ (in
  $N_{\rm v}\,r_{\rm v}^{-2}$)
\label{ilopstats} 
}
\centering
\begin{tabular}{llc}
\hline\hline
\cline{2-3}
Velocity cut & \multicolumn{1}{c}{No\ } & Yes \\ 
\hline
Arithmetic mean & 0.116 & \ \ 0.098 \\ 
Geometric mean & 0.102 & \ \ 0.085 \\ 
Standard deviation of $\log\hat\Sigma_{\rm i}$ & 0.216 & \ \ 0.229 \\ 
Spearman rank correlation &  $0.025$ & $-0.008$ \\ 
Probability & 0.34 &\,0.44 \\ 
\hline
\end{tabular}
\tablefoot{Spearman rank correlation is between mean surface density of
interlopers and halo mass.
Probability is of having a stronger correlation by chance.}
\end{table}
the arithmetic mean value of $\Sigma_{\rm i}$ matches well the value of the
stacked virial cone, regardless of the velocity cut, which has only a minor
effect on the statistics of interlopers (while the geometric means are
$\simeq 15\%$ lower).
But the dispersion in $\log \hat\Sigma_{\rm i}$ is as high as 0.22, close to
$\sigma(\log A)$ (eq.~[\ref{sigmalogs}]), so that the relative dispersion of 
$\Sigma_{\rm i}$ is as high as a factor
$10^{0.22}\simeq5/3$.

The fraction of interlopers in the stacked virial cone
is
\begin{equation}
f_{\rm i} = {N_{\rm i}\over N_{\rm h}+N_{\rm i}}=
{\pi \,\hat \Sigma_{\rm i} \over 1+\pi\,\hat \Sigma_{\rm i}}
\label{fi}
\end{equation}
(where indices `h' and `i' correspond to halo and interloper particles,
respectively), where the second equality of equation~(\ref{fi}) made use of
$\hat \Sigma_{\rm h} = \Sigma_{\rm h}/[N_{\rm v}/ r_{\rm v}^2] = 1/\pi$, by
definition. 
This yields
$f_{\rm i} = 27\%$ before and
$23.1$$\pm$$0.1\%$ after the velocity cut, where the error is both from binomial
statistics and from a bootstrap on all
the particles; a bootstrap on the halos leads to an error of 0.6\%;
propagating (with eq.~[\ref{fi}]) the error on the mean of $\hat\Sigma_{\rm i}$ (from the standard
deviation of $\hat\Sigma_{\rm i}$ given in
Table~\ref{ilopstats}) also leads to an error on $f_{\rm i}$ of 0.6\%; finally,
the
standard deviation of the interloper fraction for the three stacked cartesian
virial cones is 1.7\%.
We adopt the error estimate of 0.6\% on $f_{\rm i}$ for the later discussion.

The small decrease in interloper fraction from
before to after the velocity cut confirms our finding that the large
majority of interlopers have too low velocities to be filtered by velocity.
Cosmic variance causes huge fluctuations in the
fraction or surface density of interlopers  (2/3 of the mean value,
independent of the presence of a velocity cut),
with roughly a log-normal distribution (see the right panel of
Fig.~\ref{sdensvsm} and Table~\ref{ilopstats}). 
The last two lines of Table~\ref{ilopstats} indicate that there is no
statistically significant correlation of surface density of interlopers 
with halo mass.

\section{Biases in concentration and anisotropy?}
\label{conc}

What are the effects of the Hubble flow on estimates that observers make on
halos, e.g. the concentration and the velocity anisotropy 
of the distribution of their tracer
constituents (i.e. galaxies in clusters)?

\subsection{Effects of the Hubble flow}

We begin by a na\"{\i}ve comparison of the observable distributions with and
without the Hubble flow, before comparing the concentration and anisotropy
measured by an observer with the corresponding quantities we directly infer
in 3D
from the cosmological simulations.
Admittedly, the distribution of velocities without the Hubble flow is not fully
realistic, since the cosmological simulation solved equations for comoving
coordinates in an expanding universe.\footnote{In fact, static
universes are never simulated in a cosmological context, 
because of their lack of realism, given the expansion of the
Universe as seen in the Hubble law, and also because of the lack of knowledge
of suitable initial conditions in such a static universe.}

\begin{figure}[ht]
\centering
\includegraphics[width=9cm]{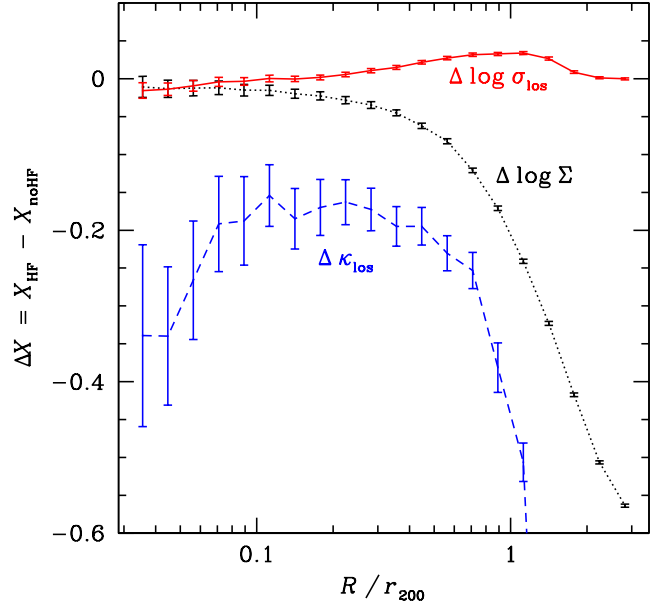}
\caption{Difference of velocity moments with Hubble flow and velocity
  cut (HF) and without Hubble flow or velocity cut (noHF):
log surface density (\emph{dotted black}),
log line-of-sight velocity dispersion (\emph{solid red}),
and
line-of-sight velocity kurtosis (\emph{dashed blue}).
The error bars are based upon Poisson errors for the surface density and
bootstraps within the radial bin for
the log dispersion and the kurtosis, where the error on the
difference is 
the square root of the sum of the square errors.
\label{hbias}}
\end{figure}
Figure \ref{hbias} shows the changes in the radial profiles of
surface density and line-of-sight
velocity dispersion and kurtosis,\footnote{The 
reader should not confuse the line-of-sight velocity
  kurtosis $\kappa_{\rm los}$ with the velocity cutoff in units of
  line-of-sight velocity dispersion ($\kappa$) or of virial velocity
  ($\hat\kappa$).} 
once the Hubble flow is added to the
peculiar velocities.
One striking feature of Fig.~\ref{hbias} is that 
the Hubble flow leads to a lower surface density
profile at large radii.
This cannot be a consequence of the restriction of the line-of-sight of the
\emph{halo} component to $\pm$$19\,r_{200}$ (see Sect.~\ref{intro}),
because the NFW surface density with line-of-sight limited to the sphere of that radius
(appendix~\ref{appnfwsph}) matches the NFW surface density projected to
infinity to better than 
1.4\% relative accuracy for $R < 3\,r_{200}$ (for $c=4$). 
Instead, it is the integral along the line of sight
of the \emph{constant density} (foreground/background) \emph{component} 
that diverges when no
Hubble flow is present, and is limited to half a box size here: $L/2=96 \,
h^{-1} \, \rm Mpc$, which corresponds to roughly 100 virial radii.
In any event, the lower (40\% lower at $r_{200}$) 
surface density profile found when  the Hubble flow
is added to the peculiar velocities
might explain the lack of concavity in the (log-log)
surface density profile within the virial radius (Fig.~\ref{sdens}).

\begin{table*}[hbt] 
\caption{MLE fits to the distribution of projected radii of the three cartesian
  stacked cones
\label{c2dfits}}
\centering
\tabcolsep 5pt
\begin{tabular}{llcclcclcccl}
\hline\hline
Model & $R_{\rm max}$ & vel. cut & \multicolumn{2}{c}{no bg} & &
\multicolumn{2}{c}{fixed bg} & & \multicolumn{3}{c}{free bg} \\ 
\cline{4-5}
\cline{7-8}
\cline{10-12}
& & & $c_{\rm 2D}$ & $P_{\rm KS}$ & &  $c_{\rm 2D}$ & $P_{\rm KS}$ & &
$100\,\hat\Sigma_{\rm bg}$ \ ($\sigma(\log\hat\Sigma_{\rm bg})$) & $c_{\rm
  2D}$ & $P_{\rm KS}$ \\
(1) & (2) & (3) & (4) & (5) & & (6) & (7) & & (8) & (9) & (10)\\
\hline
NFW    &1&N&3.46$\pm$0.04$\pm$0.21&  $<$$10^{-5}$&&4.06$\pm$0.05$\pm$0.26&      0.18&&2.4 \ (0.20)&3.95$\pm$0.05$\pm$0.03&      0.08\\ 
NFW    &1&Y&3.84$\pm$0.05$\pm$0.19&     0.059&&4.10$\pm$0.05$\pm$0.21&      0.19&&1.0 \ (0.07)&4.06$\pm$0.11$\pm$0.17&       0.2\\ 
Einasto&1&N&3.30$\pm$0.04$\pm$0.21&   0.00053&&3.90$\pm$0.05$\pm$0.25&      0.46&&2.6 \ (0.18)&3.83$\pm$0.05$\pm$0.08&      0.27\\ 
Einasto&1&Y&3.70$\pm$0.04$\pm$0.18&      0.32&&3.96$\pm$0.05$\pm$0.20&      0.42&&0.8 \ (0.01)&3.87$\pm$0.12$\pm$0.19&      0.65\\ 
NFW    &1.35&N&3.21$\pm$0.03$\pm$0.24&   $<$$10^{-14}$&&4.07$\pm$0.05$\pm$0.31&      0.11&&2.9 \ (0.08)&4.06$\pm$0.07$\pm$0.15&      0.25\\ 
NFW    &1.35&Y&3.78$\pm$0.04$\pm$0.23&     0.052&&4.19$\pm$0.05$\pm$0.26&      0.06&&0.5 \ (0.29)&3.96$\pm$0.07$\pm$0.16&      0.22\\ 
Einasto&1.35&N&3.00$\pm$0.03$\pm$0.22&   $<$$10^{-13}$&&3.83$\pm$0.04$\pm$0.28&      0.32&&3.3 \ (0.08)&3.95$\pm$0.06$\pm$0.16&      0.52\\ 
Einasto&1.35&Y&3.59$\pm$0.04$\pm$0.20&      0.04&&3.98$\pm$0.04$\pm$0.22&      0.46&&0.9 \ (0.10)&3.89$\pm$0.08$\pm$0.17&      0.73\\ 
NFW    &3&N&1.67$\pm$0.01$\pm$0.16&         0&&3.82$\pm$0.04$\pm$0.37&         0&&3.8 \ (0.03)&4.42$\pm$0.04$\pm$0.45& $<$$10^{-6}$\\ 
NFW    &3&Y&3.06$\pm$0.02$\pm$0.12&         0&&4.31$\pm$0.04$\pm$0.17&  $<$$10^{-9}$&&1.2 \ (0.12)&4.34$\pm$0.05$\pm$0.38& 0.00003\\ 
Einasto&3&N&1.42$\pm$0.01$\pm$0.12&         0&&3.19$\pm$0.03$\pm$0.34&         0&&4.3 \ (0.05)&4.15$\pm$0.03$\pm$0.67&  0.00003\\ 
Einasto&3&Y&2.65$\pm$0.02$\pm$0.10&         0&&3.75$\pm$0.03$\pm$0.12&  $<$$10^{-13}$&&1.5 \ (0.03)&4.02$\pm$0.01$\pm$0.12&   0.00055\\ 
\hline
\end{tabular}
\tablefoot{Column 1: model (NFW or $m$=5 Einasto); 
column 2 ($R_{\rm max}$): projected radius of the cone in which the stacked
cluster is built, in units of $r_{200}$;
column 3 ($v$-cut): presence (Y) or absence of the velocity cut with
$\kappa=2.7$ (NFW) or 2.6 (Einasto);
columns 4--5, 6--7, 9--10: mean best-fit concentration  ($c=r_{200}/r_{-2}$) from
projected radii and probability that distribution of projected radii is
consistent with model using a Kolmogorov-Smirnov test ($P_{\rm KS}$), for
fits without a background (cols. 4 and 5), with a fixed background
($\hat\Sigma_{\rm bg} = 0.0286$ [no velocity cut] or 0.0126 [with velocity cut], cols. 6 and 7) or a free
background (cols. 9 and 10), with best-fit value (100 times the geometric mean and error on its
logarithm in parentheses) given in column 8. 
The minimum projected radius is set to $0.03\,r_{200}$.
For the Einasto model, we adopt
the approximation to the surface density and projected number (mass) profiles
given in appendix~\ref{appsdensEinasto}. 
The errors
on $c$ are statistical (first) and a measure of the cosmic variance
term estimated by the gapper (\citealp{WT76}, see
\citealp{BFG90}) standard deviation of the MLE values for the three projection axes.}


\end{table*}

We noticed in Sect.~\ref{removal} that the line-of-sight velocity dispersion
profile showed an excess at radii near the virial radius.
Figure~\ref{hbias} confirms that the line-of-sight velocity dispersion
profile is gradually overestimated at large radii, while the surface density
profile is 
much more biased beyond half a virial radius, being underestimated
at large radii. 
The effects of the Hubble flow on the surface density and line-of-sight
velocity dispersion profile are both small at very low projected radii.

Finally,  our sharp cut (Sect.~\ref{removal})
in the 
distribution of line-of-sight velocities when the Hubble flow is incorporated
implies that 
\emph{the line-of-sight velocity kurtosis is underestimated (especially at
large projected radii)}.

\subsection{Concentration}

Does the excess of interlopers at large projected radii lead to lower values
of the concentration parameter in the fits of the projected NFW and Einasto profiles to the
surface density profiles of clusters?

Table~\ref{c2dfits} shows the MLE fits  (see
appendix~\ref{appmle}) of 
the NFW and $m$=5 Einasto surface density profiles to the distribution of projected
radii of the three cartesian stacked cones. Different fits were performed 
with variations in
the maximum allowed projected radius, $R_{\rm max}$,
the presence of a constant (fixed or free) background term, and the possible removal of
high-velocity outliers.
The NFW surface density and projected number (or equivalently projected
mass) profiles, required for the normalization of the probability used in
the MLE, are given by
\cite{Bartelmann96} and, in another form by \cite{LM01}.
The surface density and projected number (mass) profiles of the Einasto model
are not known in analytical 
form, so we have derived accurate approximations in
appendix~\ref{appsdensEinasto}. For these MLE, we adopt
equation~(\ref{Sigmaapx2})  with equation~(\ref{muapx}) for the surface
density profile and
equation~(\ref{Mprojapx}) with equation~(\ref{muapx}) for the projected mass profile.  
When a constant background term is included in the fits, it is either free
(column 8) or fixed at $\hat\Sigma_{\rm bg} = \hat\kappa\, B = 0.0286$ and
0.0126 without ($\hat\kappa = 4$) and with ($\hat\kappa = \eta \,\kappa = 1.75$)
the velocity cut (see Sect.~\ref{ilopstatsaftervcut}). 

The concentrations measured on the projected radii with single component fits 
are always smaller than the characteristic value found in 3D ($c=4.0$, 4.1,
and 4.5, see bold values in Table~\ref{c3d}).
The best-fit concentrations
are very low when the fits are
performed out to $3\,r_{200}$ (unless a velocity cut is performed  or
background component added to the model). 
Note that when the background is fitted together with the concentration and
no velocity cut is performed, the
best-fit value for the background can be over a factor two off from the value expected from
equation~(\ref{Sigmabg}), even with maximum projected radii of $3\,r_{200}$.
The KS tests indicate that for most combinations of maximum projected radius,
presence or absence of the velocity cut and how the background is
handled, 
the $m$=5 Einasto model usually provides a better
representation of the distribution of projected radii than does the NFW
model.
Finally, the errors in Table~\ref{c2dfits} indicate that the cosmic variance of stacks
of 93 clusters, measured using the standard deviation of the three cartesian
stacked cones with the gapper\footnote{The gapper dispersion of a vector
  $\vec x$ of length $n$ is $s =
  \sqrt{\pi}/[n(n-1)]\,\sum_{i=1}^{n-1} i (n-i) \,(x_{i+1}-x_i)$ \citep{WT76}.} estimate of
dispersion, which is most robust to small sample sizes \citep{BFG90},  are much greater than the intrinsic fitting errors.

\begin{table*}
\centering
\caption{Concentration bias of 2D fits
\label{cbiastab}}
\begin{tabular}{llccccccc}
\hline\hline
Model & \multicolumn{1}{c}{$R_{\rm max}$} & \multicolumn{7}{c}{$c_{\rm
    2D}/c_{\rm 3D}^{\rm best}$} \\
\cline{3-9}
& & \multicolumn{3}{c}{no $v$-cut} & & \multicolumn{3}{c}{$v$-cut} \\
\cline{3-5}
\cline{7-9}
& & no bg & fixed bg & free bg & & no bg & fixed bg & free bg \\
\hline
NFW      & 1 & 0.86$\pm$0.07 &{\bf 1.01$\pm$0.08} & {\bf0.99$\pm$0.05} & 
& {\bf0.96$\pm$0.06} & {\bf1.02$\pm$0.07} & {\bf1.01$\pm$0.07}\\ 
Einasto $m$=5 & 1 & \blueit{0.82$\pm$0.06} & {\bf0.97$\pm$0.08} & {\bf0.96$\pm$0.05} & 
& {\bf0.93$\pm$0.06} & {\bf0.99$\pm$0.07} & {\bf0.97$\pm$0.07}\\ 
NFW      & 1.35 & 0.78$\pm$0.06 & {\bf0.98$\pm$0.08} & {\bf0.98$\pm$0.05} & 
& {\bf0.91$\pm$0.06} & {\bf1.01$\pm$0.07} & {\bf0.96$\pm$0.05}\\ 
Einasto $m$=5 & 1.35 & 0.72$\pm$0.06 & {\bf0.93$\pm$0.07} & {\bf0.95$\pm$0.05} & 
& \blueit{0.87$\pm$0.06} & {\bf0.96$\pm$0.06} & {\bf0.94$\pm$0.05}\\ 
NFW      & 3 & 0.37$\pm$0.04 & 0.85$\pm$0.09 & 0.98$\pm$0.11 & 
& 0.68$\pm$0.04 & 0.96$\pm$0.06 & 0.96$\pm$0.10\\ 
Einasto $m$=5 & 3 & 0.32$\pm$0.03 & 0.71$\pm$0.08 & \blueit{0.92$\pm$0.16} & 
& 0.59$\pm$0.04 & 0.83$\pm$0.05 & \blueit{0.89$\pm$0.05}\\ 
\hline
\end{tabular}
\tablefoot{The biases highlighted in bold (respectively blue italics) 
show the cases
where the best-fit surface 
density profile was consistent with the data to better than 5\% (between
0.01\% and 5\%) confidence
(see last column of Table~\ref{c2dfits}). 
The errors are the statistical (MLE fit) and cosmic variance
(from the 3 cartesian stacked halos) errors of
Table~\ref{c2dfits} added in quadrature together and
with the analogous errors on the best-fit (bold in Table~\ref{c3d}) 3D
concentration, with $\sigma^2(c_{\rm 
  2D}/c_{\rm 3D}^{\rm best})=\sigma^2(c_{\rm 2D})/\left\langle c_{\rm 3D}^{\rm best}\right\rangle^2
+ \sigma^2(c_{\rm 3D}^{\rm best}) \left \langle c_{\rm 2D}\right\rangle^2/\left\langle
c_{\rm 3D}^{\rm best}\right \rangle^4$.}
\end{table*}
Table~\ref{cbiastab} shows the bias in concentrations, where the bias is the
ratio between the
concentration measured with the 2D fit (Table~\ref{c2dfits}) and the best-fitting of the
concentrations found with the 3D fits  (highlighted in bold in
Table~\ref{c3d}), for given $r_{\rm 
  max}/r_{200}$ and different models and backgrounds.
The concentration parameters found in the
fits of the surface density profile are 
underestimated by typically
16\%$\pm$7\%
when we make no velocity cut, limit the projected radii to $r_{200}$, and do
not incorporate a constant background in the fit. This underestimate of the
concentration is statistically significant (marginally so for NFW).
The concentration bias gets worse as we increase the maximum projected
radius: the concentration is underestimated by 1/4 at $R<r_{100}$ and by as
much as 2/3 with $R < 3\,r_{200}$.

But when we make the
velocity cut at 2.7 (NFW) or 2.6 (Einasto) $\sigma_{\rm los}(R)$, where $\sigma_{\rm los}(R)$ is
obtained from equations~(\ref{siglosapx}) and (\ref{siglosconvert}), 
the biases in the concentration parameters are typically reduced by a factor
two.
When we limit the analysis at $R < r_{200}$,
\emph{the concentration parameters
found in the fits of the surface density profiles are
(only) underestimated by typically
6$\pm$6\%}.
This low bias is no longer statistically significant given our fit and
cosmic variance errors.\footnote{We also experimented with concentration fits
  on the projected radii of all particles within $3\,\sigma_{\rm los}(R)$
  instead of 2.7, but the changes 
were very small (less than 1\%, with slightly worse underestimates of the
concentration for the single component fits).} 
However, extending the analysis to $R = r_{100}$, the concentration is still
biased low by as much
as 11\% (which is
marginally significant), despite the velocity cut. And if we go all the way to
$3\,r_{200}$, the bias is still very strong, as the concentration is
underestimated by over 1/3.

One may wonder whether one can recover the concentration parameter more
accurately with two-component fits to the set of projected radii than with
the single-component fit, since we found an
additional background needs to be added (Sect.~\ref{statswohivilop}).
For example, \cite{LMS04} fit a projected NFW model plus a constant surface
density background (hereafter NFW+bg), with no velocity cut, for projected
radii $0.02 < R_{\rm
  max}/r_{200} < 2.5$.
We tested the single and double-component models 
in the optimistic case of our stacked cluster with nearly
$3\times10^5$
particles. 
As seen in Table~\ref{cbiastab}, \emph{the 3D concentration parameter
is recovered better by the two-component models,
regardless of the velocity cut},
although the improvement is not always statistically significant.
Note that the background  is well recovered with $R_{\rm max} = 3\,r_{200}$
and a velocity cut for both the NFW+bg and
Einasto+bg models (Table~\ref{c2dfits}).

In summary, \emph{the concentrations measured in 2D recover best the 3D
  values once the velocities are filtered and especially once a background
  is included in the fit (even when limited to the virial radius)}.

Note also that 
one should not attempt to model the surface density profile as the sum of the
halo term (with \emph{line-of-sight limited to the sphere}, with the formulae of
appendix~\ref{appnfwsph} for the NFW model) 
and a constant background,
because the \emph{total} surface density profile decreases smoothly beyond
the virial radius in ways that are not simple to model, for example with a
spherical halo and a constant background.
Moreover, for fits where the projected radii are limited to the virial
radius, these spherical plus background  fits are not recommended because the
background is not constant but rises with radius (Figs.~\ref{Rhists} and
\ref{univvsrad}) and is less easy to model than the surface density with
line-of-sight integrated to infinity.

\subsection{Velocity anisotropy}

The signature of the Hubble flow on the shape of the line-of-sight velocity
dispersion profile (Fig.~\ref{siglos3}) suggests that the velocity anisotropy
that 
is recovered may be affected.
\begin{figure}[ht]
\centering
\includegraphics[width=9cm]{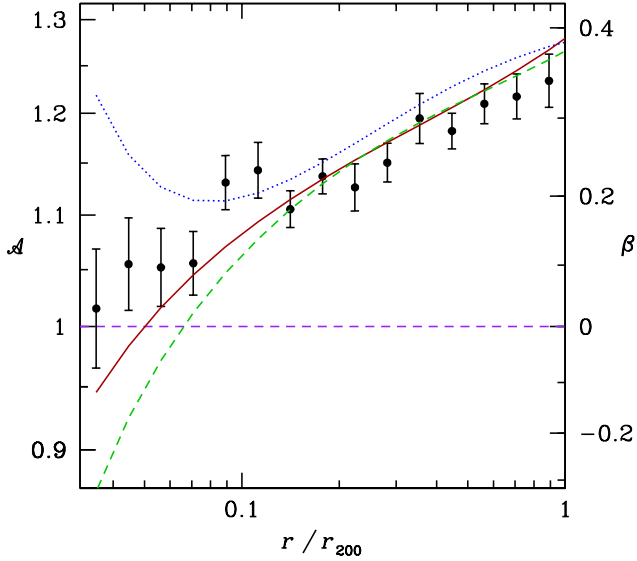}
\caption{Velocity anisotropy profiles (including streaming motions:
  eqs.~[\ref{betadef}] and [\ref{betapdef}]) of the stack of the 93 halos. 
The \emph{points} are the measured velocity anisotropy 
(same as in Fig.~\ref{betafig}, again with uncertainties from 100 bootstraps
on the  93 halos) 
and the \emph{curves} are recovered from anisotropy inversion assuming the 
  $c$=4 NFW model
  (\emph{solid curves}), for three polynomial fits (orders 2, 3, and 4 in
log-log space) to the  
measured line-of-sight velocity dispersion profile (after the  $\kappa=2.7$ 
velocity cut using the NFW model with ML anisotropy: \emph{solid dark red},
\emph{dashed green},
and \emph{dotted blue} for orders 2 to 4, respectively).
The \emph{purple dashed horizontal line} indicates the fully isotropic case.
Note that the 4th order polynomial (\emph{blue}) extrapolates poorly the line-of-sight
velocity dispersion profile at very low and very
high projected radii.
\label{betainv}
}
\end{figure}
Figure~\ref{betainv} shows the result of the non-parametric anisotropy
inversion (first 
developed by \citealp{BM82}, but we use here the simpler algorithm by
\citealp{SS90}), which computes the anisotropy profile assuming a smooth
representation 
of the
line-of-sight velocity dispersion profile and a mass model.
Here, we adopt
 a $c$=4 NFW model and fit polynomials to the binned $\log \sigma_{\rm los}$
 vs. $\log R$.\footnote{We cannot employ the analytical approximation of
   equation~(\ref{siglosapx}) to the line-of-sight velocity dispersion profile for an
   NFW model with $r_{\rm a} = r_{-2}$ ML anisotropy, because we place
   ourselves in the context of an observer who wishes to measure the velocity
   anisotropy with no prior on it: (s)he is thus forced to use a smooth
   representation of the observed line-of-sight velocity dispersion profile.}

In the region where the order of the polynomial fit does not matter ($0.1 <
r/r_{200}<1$), 
the recovered anisotropy profile reproduces very well the one measured in
three dimensions (points in Fig.~\ref{betainv}), although, beyond $0.2\,r_{200}$, 
the recovered anisotropy profiles are slightly more radial than that
measured in 3D.
This bias towards more radial motions appears statistically significant,
since in all six radial bins where there is a $\simeq 1\,\sigma$ offset, this
offset is in the same direction (probability of $2^{-5}=3\%$).
Therefore, \emph{the Hubble flow produces only a slight radial velocity
  anisotropy bias in the envelopes of halos.}

\section{Summary \& discussion}
\label{discuss}

This work analyzes the distribution of particles in projected phase space
around dark matter halos in cosmological simulations. The particles are split
among halo particles within the virial sphere and interlopers within the
virial cone but outside the virial sphere  (Fig.~\ref{scheme}). 
The reader should be careful that 
the analyses presented here cannot be directly applied to observations of 
clusters of galaxies, as they work with halo particles instead of galaxies within
clusters, and assume the halo centers to be determined quite precisely (from
real space measurements).

We find a universal distribution of interlopers in projected
phase space, i.e. with little dependence on halo mass (Figs.~\ref{vhistint1}c
and \ref{univvsmass}).
In particular, we note that velocity cuts cannot
distinguish the quarter of particles that are interlopers from those in the
virial sphere (Fig.~\ref{ilopfracs}), as was previously noted by \cite{Cen97}.
We find that the distribution of interlopers in projected phase space displays
a roughly constant 
surface density (Figs.~\ref{Rhists} and \ref{univvsrad})
 and a  distribution of line-of-sight 
velocities that is the sum of a quasi-gaussian component, caused by the halo
outskirts (out to typically 8 virial radii, Fig.~\ref{vhistint1}b) 
and a uniform component caused by
particles at further distances from the halo (Figs.~\ref{vhists} and
\ref{vhistint1}).

The cosmological simulations allow us to optimize the ratio of maximum
velocity to line-of-sight velocity dispersion that recovers the latter
quantity. Although this may seem to be a circular argument (since
$\sigma_{\rm los}(R)$ depends on the velocity cut), it has been
widely used in the past, usually in iterative form, with a $3\,\sigma$
cutoff. We find that this cutoff is not restrictive enough and causes an
overestimate of the line-of-sight velocity dispersion profile (based upon
mass and velocity anisotropy models derived from the cosmological
simulations): up to 10\% for the isotropic NFW velocity cut, which is reduced
to 5\% for the ML anisotropy velocity cut (Fig.~\ref{siglos3}).
We recommend
instead a velocity cut at $2.7 \,\sigma_{\rm los}(R)$ on the best iterative
fit to the line-of-sight velocity dispersion for the NFW model with
$r_{\rm a}=r_{-2}$ ML anisotropy. Alternatively, one can use a velocity
cut at $\kappa = 2.6$ for the $m$=5 Einasto model, modeled (again) with $r_{\rm a}
= r_{-2}$ ML anisotropy, but this underestimates the line-of-sight velocity
dispersion near the virial radius
(Fig.~\ref{siglos3}).

We illustrate (Figs.~\ref{phasespacewcuts}, 
\ref{Rvzdist}, \ref{vhists}, \ref{ilopfracs}, and \ref{sdens}) 
how the distribution of particles in projected phase space is
altered once the high velocity interlopers are rejected with this new
velocity filter (besides limiting the line-of-sight to typically $\pm
17\,r_{200}$, the main effect is to remove the flat velocity component).
The fraction of interlopers within the virial cone drops from 27\% (with an
observer at distance $D=90 \, h^{-1} \, \rm Mpc$) to $23.1$$\pm$$0.6\%$
(independent of $D$ for $D\ga 17\,\left\langle r_{200}\right\rangle$)
when the velocity cut is applied (where the uncertainty is taken from the end
of Sect.~\ref{ilopstatsaftervcut}).

This fraction of interlopers can be directly inferred from the NFW or Einasto
model
\begin{equation}
f_{\rm i} = {\hat M_{\rm p}(r_{200})-1 + \pi\,\hat\Sigma_{\rm bg} \over
\hat M_{\rm p}(r_{200}) + \pi\,\hat\Sigma_{\rm bg}} \,
\label{fofi}
\end{equation}
where $\hat M_{\rm p} = M_p/M_{200}$ is the projected virial mass in virial
units (i.e. in units of the mass within the virial sphere), while
$\hat\Sigma_{\rm bg}$ is given in equation~(\ref{Sigmabgvir}).
For the NFW model, one then obtains $f_{\rm i}$ = 26.8\% and 24.0\%,
respectively before ($\hat\Sigma_{\rm bg}=0.0286$) and after  ($\hat\Sigma_{\rm 
  bg}=0.0126$) the velocity cut, while with the $m$=5 Einasto model with
$c=4$, the corresponding percentages of interlopers are 25.7\% and 22.8\%.
These theoretical predictions are
in excellent agreement with the fractions
obtained from the simulations.
Note that the omission of the
background ($\hat\Sigma_{\rm bg}$) term in equation~(\ref{fofi})  reduces
$f_{\rm i}$ by typically 10\% in relative terms, relative to the fractions
after the velocity cut. 
Applying equation~(\ref{fofi})
to models of different concentrations leads to roughly a power-law variation
of $f_{\rm i}$ with slope --0.32 (NFW) or --0.49 ($m$=5 Einasto).
Therefore, \emph{the fraction of interlopers should be (slightly) more
  important in the more 
massive halos}, since they have (slightly) lower concentrations
\citep{NFW97,MDvdB08}.

In comparison, using 62 clusters from the same simulation as the one we have
analyzed (we have 53 clusters in common), 
\cite{Biviano+06} found that among particles selected in cones of projected
radius $1.5 \, h^{-1} \, \rm Mpc$ around cluster-mass halos (after their velocity cut),
$18$$\pm$$1.4$\% of them lie outside the sphere of the same radius.\footnote{The error is
  taken as 
their dispersion over the square
root of their number of halos.\label{error}}
Their halos have a median virial radius of $r_{200} = 0.93 \, h^{-1} \, \rm
Mpc$ (1.08 times our median) 
and hence a virial mass of $M_{200} = 1.9\times 10^{14} h^{-1} M_\odot$.
Their Figure~7 indicates that their velocity cut is roughly 1180, 1105, and $780\,\,
\rm km \, s^{-1}$, at projected radii 0.6, 1.0 and $1.5 \, h^{-1} \, \rm
Mpc$, respectively. Since NFW concentration scales as $M^{-0.1}$
\citep{NFW97,MDvdB08}, their median concentration should be 3.9, hence their scale radius should be 
$930/3.9 = 238 \, h^{-1} \, \rm kpc$. 
Assuming an NFW model, we deduce that their median circular velocity at the
scale radius is $914 \, \rm km \, s^{-1}$, and find that their velocity
cuts correspond to $\kappa = 2.1, 2.2$, and 1.8, at the three projected radii
chosen above. These fractions are consistent with the values of $\hat\kappa$
one can read off of Figure~3 of \cite{Wojtak+07} that illustrates the same
velocity cut model \citep{dHK96}.
We then considered a cone of projected size $1.5/0.93 =
1.6\,r_{200}$. Adopting their typical $\kappa = 2$, we then
found that after a $2\,\sigma$ velocity cut, 
the fraction of particles with $r > 1.6\,r_{200}$ is now 21.3\%.
This fraction is still marginally significantly larger than
\citeauthor{Biviano+06}'s
fraction of 18\%
 (assuming the
same errors as above).
We attribute this discrepancy to their variable $\kappa$ velocity cut, 
which differs from our fixed $\kappa$ one.
\cite{Wojtak+07} tried several interloper removal schemes and definitions
(using a different $\Lambda$CDM cosmological simulation). Their
local $3\,\sigma$ cut leads to $20.4$$\pm$$1.7$\% of interlopers remaining within the
virial cone.
Given the quoted errors, the lower fraction of
interlopers found by \citeauthor{Wojtak+07} is marginally consistent with ours.

\emph{This fraction of 23\% of interlopers after the velocity cut
is surprisingly close to the
fraction of blue galaxies} (i.e. galaxies off the Red Sequence) observed within SDSS
clusters, as \cite{YMvdB08} find roughly 22\% of blue galaxies within
SDSS clusters  of masses $>10^{14} h^{-1} M_\odot$. 
Admittedly, it is dangerous to match the dark matter  distribution with the
galaxy distribution, since galaxies  are
biased tracers of the matter distribution. In fact, galaxies are biased
relative to dark matter halos (e.g. \citealp{CWK06}), which in turn are
biased relative to the dark 
matter particle distribution (e.g. \citealp{MW96,CLMP98}).
If, in the end, the SDSS galaxies analyzed by \citeauthor{YMvdB08} are
unbiased tracers of the dark matter distribution, then
this close agreement would be 
expected if all blue galaxies are caused by projection effects. But if
projections also pick up red galaxies in groups, then some blue galaxies
would need to survive within the virial sphere for the match to hold.
However, the \citeauthor{YMvdB08} group finder is fairly
efficient in separating groups along the line-of-sight, so we conclude that
the fraction of
blue galaxies within the virial sphere should be small. In other words, star
formation appears to be strongly quenched when galaxies penetrate the virial
spheres of clusters.

When no velocity cut is performed, 
a maximum likelihood fit of the concentration of
the projected NFW model to the projected radii of a stacked cluster of nearly
$300\,000$
 particles out to $r_{200}$ ($r_{100}$)
leads to a 14$\pm$7\% (22$\pm$6\%) underestimate of the true concentration
 parameter
(Table~\ref{cbiastab}, where most of the uncertainty comes from cosmic
 variance).
Similar biases occur with the $m$=5 Einasto model.
But after the velocity cut, these biases decrease by a factor two, and are no longer
statistically significant (Table~\ref{cbiastab}). 
Moreover, the inclusion in such fits of a constant background as an extra parameter 
also strongly decreases the bias, even
when the maximum projected radius is as low as 
$r_{200}$
(Table~\ref{cbiastab}). 
In fact, inspection of Table~\ref{cbiastab} indicates that, for $R_{\rm max}
= r_{100}$ or $3\,r_{200}$,
the background (fixed or free) has a greater influence than the velocity
filter in removing the bias on measured concentration.
Surprisingly, for $R_{\rm max} = 3\,r_{200}$,
a physically motivated fixed background added to the NFW
model is slightly less effective in reducing the concentration bias than is a
free background.

When the maximum
radius is
$3\,r_{200}$ and no velocity cut is performed, 
the NFW model with a free (respectively fixed) background underestimates the
concentration (Table~\ref{cbiastab}) 
by $2$$\pm$$11\%$ ($15$$\pm$$9\%$).
This insignificant (marginally significant) bias 
is caused by the strong decrease of the surface density profile once the
Hubble flow is added to the peculiar velocities (Fig.~\ref{hbias}).
These small biases suggest that the fairly low concentration
($c_{200}=2.9$$\pm$$0.2$) for the galaxy 
distribution in clusters found by \cite{LMS04}, who fit an NFW model with a
free constant background to the distribution of projected 
radii in the range $0.02 < R/r_{200} < 2.5$, but who did not make a velocity
cut for lack of velocity data, is incompatible with true
cluster concentrations of $c=4.0$
at the $2\,\sigma$ level.
The lower concentration bias with the two-component model is expected,
because the single 
component NFW or Einasto models cannot capture the flat surface density at
large radii (Fig.~\ref{sdens}), because other halos are projected along the
line-of-sight.

While a two-component model of halo (to infinity) + constant background is
better able to recover the halo concentration than a single-component model
(Table~\ref{cbiastab}),
it is not wise to estimate the halo concentration from a two-component model
with a halo term
whose line-of-sight is limited to the sphere (Appendix~\ref{appnfwsph}) 
plus a near constant
background term arising from our universal interloper surface density model 
(eq.~[\ref{SigmaivsA}] or simply [\ref{SigmaiNFW}]): the single-component NFW
captures better the \emph{total} 
surface 
density profile than this halo+background model, 
especially if the maximum projected radius is beyond the
virial radius, as the interloper surface density has a discontinuous
slope at the virial radius (Fig.~\ref{sdens}).
On the other hand, the universal distribution of interlopers in projected
phase 
space might be
useful to model the internal kinematics (hence total mass
profile) of clusters of galaxies, where the full distribution of galaxies in
projected phase space is the sum of these interlopers and an NFW-like model
projected onto the virial sphere. We are preparing tests of the
mass/anisotropy modeling of clusters, groups, and galaxies (through their
satellites) using this interloper model.

We also performed 2D fits to individual halos of typically 700 particles (not
shown here). The dispersion of the concentrations were much larger
(typically 0.16 dex)
than the biases 
obtained from the stacked virial cone (typically 0.05 dex, i.e. 10\% errors,
see Table~\ref{cbiastab}), which means that shot noise and cosmic
variance dominate the bias caused by the Hubble flow.

The line-of-sight velocity dispersion profile shows a concavity (in log-log)
near the virial radius (Fig.~\ref{siglos3}), which is caused by the Hubble flow
(Fig,~\ref{hbias}). 
The velocity
anisotropy profile recovered from this velocity dispersion profile, assuming
the correct mass distribution, is close to the true anisotropy profile, with
a slight, marginally significant, radial bias in the envelopes of
clusters in comparison with the anisotropy profile recovered in 3D
(Fig.~\ref{betainv}), as was previously noted by \cite{Biviano07}. 

In summary, the density profile of $\Lambda$CDM halos falls fast enough that
the effects of the Hubble flow perturbing the standard projection equations
produce only small biases in comparison with the shot noise of clusters with less
than 1000 galaxies, as well as the large cosmic variance of the halos.

These results have been obtained with the dark matter particles of a
cosmological $N$-body simulation (with additional gas and galaxy
components).  They will need to be confirmed with future
more realistic simulations of the \emph{galaxy} distribution.

\begin{acknowledgements}
We thank Marisa Girardi for providing the positions and masses of the mock
clusters, Mike Hudson and Raphael Gavazzi for helpful comments,
and Richard Trilling for a critical reading of an early version of the
manuscript. 
We also warmly thank an anonymous referee for his thorough reading of the
manuscript and several important comments, especially his insistence on our
use of cluster bootstraps to estimate the errors from cosmic variance.
AB acknowledges the hospitality of the Institut d'Astrophysique de
Paris. This research has been partly financially supported by INAF through
the PRIN-INAF scheme. 
The simulation has been carried out at the Centro interuniversitario el
Nord-Est per il Calcolo Elettronico (CINECA, Bologna) with CPU time
assigned thanks to an INAF-CINECA grant.

\end{acknowledgements}

\bibliography{master}

\onecolumn
\appendix
\section{Projected mass, surface density and tangential shear of the Einasto model}
\label{appsdensEinasto}

In this appendix, we derive an approximation to 
the surface density and projected mass (or, equivalently, projected number)
profiles for the Einasto model.

\subsection{Projected mass profile}

For any density model, the projected mass is
\begin{eqnarray}
M_{\rm p}(R;m) &=& \int_0^R 2\, \pi \,S\,\Sigma(S;m)\,\d S \nonumber \\
&=& 4\,\pi \left [\int_0^R r \,\rho(r)\,\d r\,\int_0^r {S\,\d S\over
    \sqrt{r^2-S^2}}
+ \int_R^\infty r \,\rho(r)\,\d r\,\int_0^R {S\,\d S\over
    \sqrt{r^2-S^2}} \right] \nonumber \\
&=& 4\,\pi\,\left [ \int_0^R r^2 \,\rho(r)\,\d r + \int_R^\infty 
r\, \left (r-\sqrt{r^2-R^2}\right)\,\rho(r)\,\d r \right] 
\label{Mpgen}\\
&=& M_\infty - 4\,\pi\,\int_R^\infty r\,\sqrt{r^2-R^2}\,\rho(r)\,\d r \ ,
\label{Mpcv}
\end{eqnarray}
where the second equality is obtained after reversing the order of
integration. Equation~(\ref{Mpgen}) is general, while equation~(\ref{Mpcv})
is only valid for models with finite total mass $M_\infty$.
For the Einasto model of total mass $M_\infty$, the 3D mass profile is
\begin{equation}
M(r;m) = P \left[3m,2m \left ({r\over r_{-2}}\right)^{1/m} \right] \,M_\infty
\ ,
\label{MEinasto}
\end{equation}
where $P(a,x) = \gamma(a,x)/\Gamma(a)$ is the regularized incomplete gamma function.
The ratio 
\begin{equation}
\mu(R,m) = {M_{\rm p}(R;m)\over M(R;m)} \ ,
\label{mudef}
\end{equation}
determined from equations~(\ref{MEinasto}) and (\ref{Mpcv}), with
equation~(\ref{rhoE}),
varies little, as seen in the right panel of Figure~\ref{logratio}.
We fit again a two-dimensional fourth-order polynomial in $m$ and
$u=\log_{10}\left (R/r_{-2}\right)$ and
find
\begin{eqnarray}
\mu(R,m) &\simeq& \mu_{\rm apx} (u,m)
\label{muapxdef} \\
\mu_{\rm apx}(u,m) &=&
{\rm dex}
\left (
0.0001219\,m^4+0.0007400\,m^3u-0.003209\,m^3+0.002976\,m^2u^2-0.01560\,m^2u 
\right. \nonumber \\
&\mbox{}& \qquad +0.02966\,m^2+0.0003307\,m\,u^3-0.04434\,m\,u^2+0.1273\,m\,u-0.1149\,m
\nonumber \\
&\mbox{}& \qquad  \left.
 +0.001036\,u^4-0.003133\,u^3+0.1905\,u^2-0.5241\,u+0.3525
   \right ) \ .
\label{muapx}
\end{eqnarray}
In the interval $3.5\leq m \leq 6.5$ and $-2 \leq u \leq 2$,
equations~(\ref{muapxdef}) and (\ref{muapx}) are 
accurate to better than 1.5\% everywhere (0.23\% rms).

The projected mass of the Einasto model can thus be written
\begin{eqnarray}
M_{\rm p}(R;m) &\simeq& \mu_{\rm apx}(u,m)\,M(R;m) 
\label{Mprojapx1}\\
&=&  \mu_{\rm apx} \left [\log_{10} \left( {R\over
    r_{-2}}\right),m\right]\,P\left[3m,2\,m\,\left({R\over
    r_{-2}}\right)^{1/m}\right]\,M_\infty \ ,
\label{Mprojapx}
\end{eqnarray}
where again $u = \log_{10} (R/R_{-2})$.

\subsection{Surface density profile}

Inserting equation~(\ref{rhoE}) into equation~(\ref{projec}), the surface
density of the Einasto model of total mass $M$ and index $m$ is
\begin{eqnarray}
\Sigma(R;m) &=& {M\left(r_{-2}\right)\over \pi r_{-2}^2}\,\widetilde \Sigma \left ({R\over
  r_{-2}};m\right) \ ,
\label{Sigmadef} \\
\widetilde \Sigma(X;m) &=& {(2m)^{3m-1}\over \gamma(3m,2m)}\,\int_X^\infty \exp
\left (-2\,m\, x^{1/m}\right)\,{x \,\d x\over \sqrt{x^2-X^2}} \ .
\end{eqnarray}

\begin{figure}[ht]
\centering
\includegraphics[width=8cm]{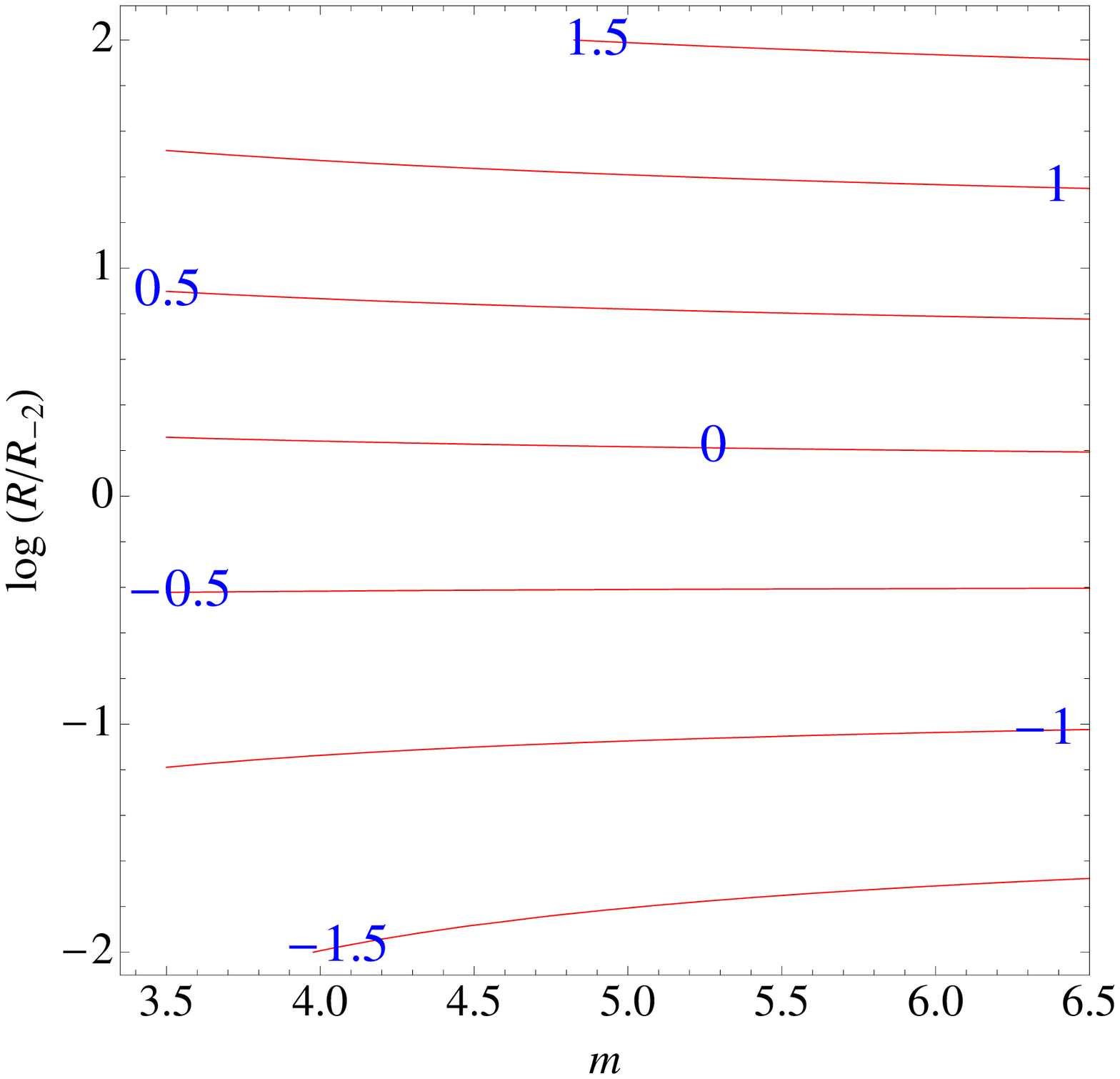}
\includegraphics[width=7.8cm]{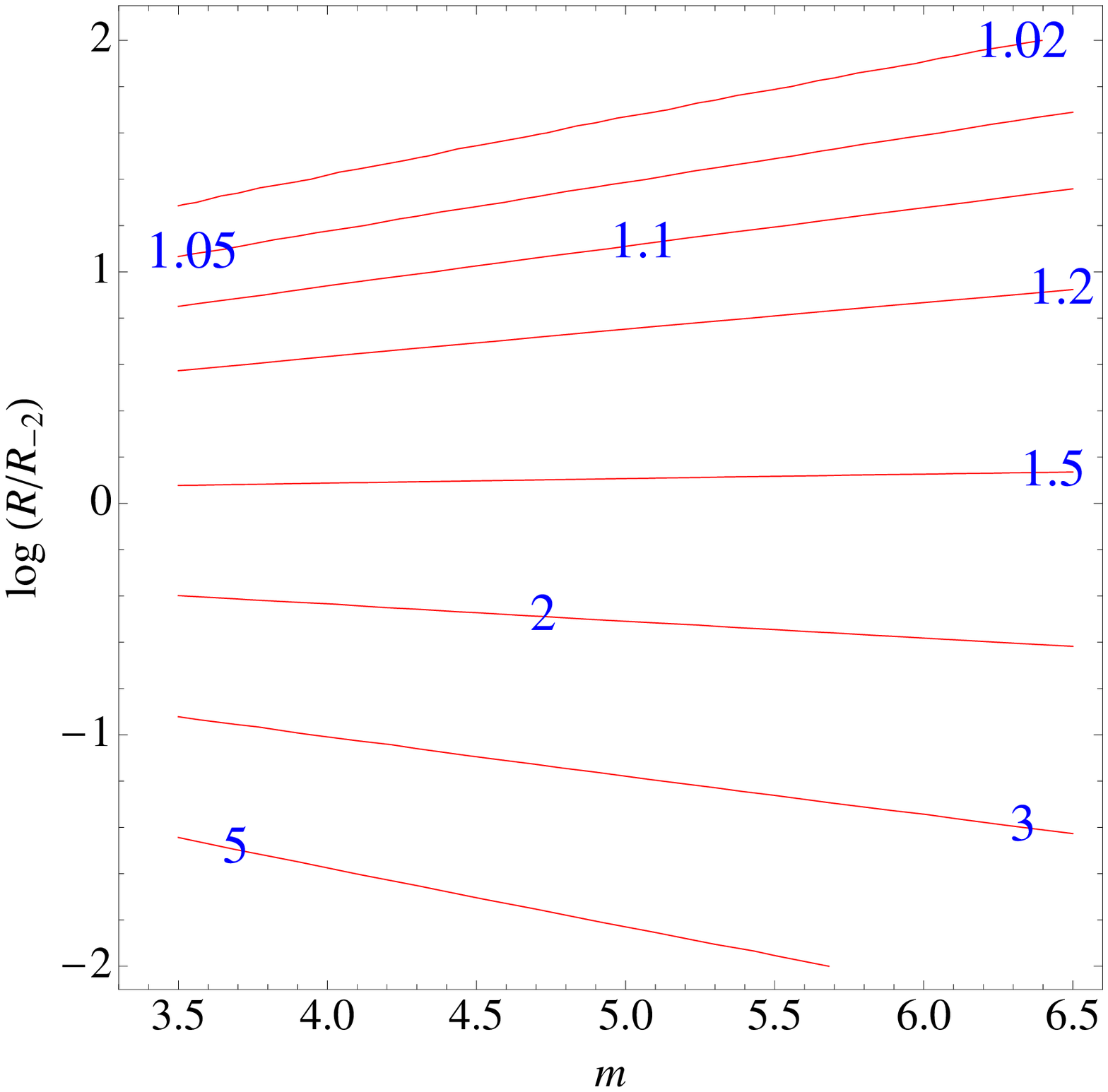}
\caption{Contours of $\log_{10} \cal R$ (\emph{left}, eq.~[\ref{calRdef}])
and
$\mu$ (\emph{right}, eq.~[\ref{mudef}], with
  eqs.~[\ref{rhoE}], [\ref{Mpcv}] and [\ref{MEinasto}]) for the Einasto model.}
\label{logratio}
\end{figure}

Writing the dimensionless mass density as
\begin{equation}
\widetilde \rho(x;m) = {\rho(x r;m) \over M\left(r_{-2};m\right) / \left( 4\pi
  r_{-2}^3\right)}
= {(2m)^{3m}\over  m\,\gamma(3m,2m)}\,\exp\left(-2m\,x^{1/m}\right)
\ ,
\label{rhotilde}
\end{equation}
where the second equality derives from equation~(\ref{rhoE}), 
we can express the ratio of dimensionless surface to space densities as
\begin{equation}
{\cal R}(X,m) = {\widetilde \Sigma(X;m) \over \widetilde
  \rho(X;m)} = 
{1\over 2 }\,\exp\left (2\,m\,X^{1/m}\right)\,
\int_X^\infty \exp\left(-2m \,x^{1/m}\right)\,{x\,\d x\over \sqrt{x^2-X^2}}
\ ,
\label{calRdef}
\end{equation}
where $X=R/r_{-2}$.
In the range $3.5 \leq m \leq 6.5$ (spanned by $\Lambda$CDM halos in the
redshift range $0 \leq z \leq 3$ according
to \citealp{Gao+08}) and $-2
\leq \log_{10} X \leq 2$,
$\cal R$ varies little and regularly, as seen in the left panel of 
Figure~\ref{logratio}.
We fit a two-dimensional 4th-order polynomial in $m$ and $u=\log_{10}
(R/r_{-2})$ to $\log_{10} {\cal R}$. We find
\begin{eqnarray}
{\cal R}(X,m) &\simeq& {\cal R}_{\rm apx} (u,m)
\label{Rapxdef} \\
{\cal R}_{\rm apx} (u,m) &=&
{\rm dex} \left (
6.286 \times 10^{-6} \,m^4+0.001178 \,m^3 u-0.0002251 \,m^3+0.001524 \,m^2 u^2-0.02427 \,m^2
   u \right. \nonumber \\
&\mbox{}& \qquad +0.0008538 \,m^2+0.001861\,m \,u^3-0.02323 \,m\,u^2+0.1849 \,m \,u+0.01577 \,m
\nonumber \\
&\mbox{}& \qquad \left. +0.0006014 \,u^4-0.01506 \,u^3+0.1056\,u^2+0.3406 \,u-0.2515 \right) \ .
\label{Rapx}
\end{eqnarray}
In the interval $3.5\leq m \leq 6.5$ and $-2 \leq u \leq 2$,
equations~(\ref{Rapxdef}) and (\ref{Rapx}) are 
accurate to better than 0.8\% everywhere (0.12\% rms).

The dimensionless surface density can then be written as
\begin{equation}
\widetilde \Sigma (X;m) = {(2\,m)^{3m}\over m\,\gamma(3\,m,2\,m)}\,\exp\left
         (-2\,m\,X^{1/m}\right)\,{\cal R}_{\rm apx}(\log_{10} X,m) \ ,
\label{Sigmaapx}
\end{equation}
or equivalently, with $M\left (r_{-2}\right) = P(3m,2m)\,M_\infty$,
where $P(a,x) = \gamma(a,x)/\Gamma(a)$ is the regularized incomplete gamma
function and $M_\infty$ the total mass of the Einasto model:
\begin{equation}
{\Sigma(R)\over M_\infty / \left (\pi r_{-2}^2 \right)} =
{(2\,m)^{3m}\over m\,\Gamma(3\,m)}\,\exp\left
         (-2\,m\,X^{1/m}\right)\,{\cal R}_{\rm apx}(\log_{10} X,m)  \ .
\end{equation}

Alternatively, 
the surface density profile can be, self-consistently, estimated from
equation~(\ref{Mprojapx1})  by
differentiation over the projected mass profile, yielding after some algebra
\begin{eqnarray} 
\Sigma(R;m) &=& {1\over 2\pi R}\,{\d  M_{\rm p}(R;m)\over \d R} \nonumber
\\
&\simeq&
{\mu_{\rm apx}(u,m) \over 2\pi R^2}\,\left [4\pi R^3 \,\rho(R;m)+ {\d \log_{10} \mu_{\rm
      apx}\over \d u}\,M(R;m) \right] \nonumber \\
&=& \left [ {M\left(r_{-2}\right)\over \pi r_{-2}^2}\right]
\,
\left [X^3\,\widetilde\rho(X;m)+{\d \log\mu_{\rm apx}\over \d \log X}
\,{P\left(3m,2m\,X^{1/m}\right)\over P(3m,2m)} \right]
\,{\mu_{\rm apx}(u,m) \over 2\,X^2} \ ,
\label{Sigmaapx2}
\end{eqnarray}
where $\widetilde \rho$ is given in equation~(\ref{rhotilde}). 

Equation~(\ref{Sigmaapx2}) has the advantage of providing an approximation
for the surface density profile that is consistent with that of the projected
mass profile. This is crucial for maximum likelihood estimation of
concentration (and possibly Einasto index and background level). On the other
hand, the accuracy of equation~(\ref{Sigmaapx2}) is about 5 times worse than
that of equation~(\ref{Sigmaapx}).   

\subsection{Tangential shear profile}

For any density model, the tangential shear measured by weak lensing can be
written (e.g. \citealp{MiraldaEscude91})
\begin{equation}
\gamma_{\rm t}(R;m) = {\overline \Sigma(R;m) - \Sigma(R;m) \over \Sigma_{\rm crit}} \ ,
\label{sheardef}
\end{equation}
where $\overline\Sigma(R;m) = M_{\rm p}(R;m)/(\pi R^2)$ is the mean surface
density, while $\Sigma_{\rm crit}=c^2/(4\pi G)\,D_{\rm S}/(D_{\rm L} D_{\rm
  LS})$ is the critical surface density, with $c$ the velocity of light, and where 
$D_{\rm S}$,
$D_{\rm L}$,
and
$D_{\rm LS}$ are the angular diameter distances between the observer and the
source, the observer and the lens, and the lens and the source,
respectively. 
Equation~(\ref{sheardef}) indicates that adding a
constant term to the surface density (eq.~[\ref{Sigmabg}])  has no effect on
$\gamma_{\rm t}$ (this is the mass-sheet degeneracy).
For the Einasto model, the tangential shear (eq.~[\ref{sheardef}]) 
is readily computed using
equations~(\ref{Sigmaapx}) with (\ref{Rapx}) and (\ref{Mprojapx}) with
(\ref{muapx}).
\begin{figure}[ht]
\centering
\includegraphics[width=9cm]{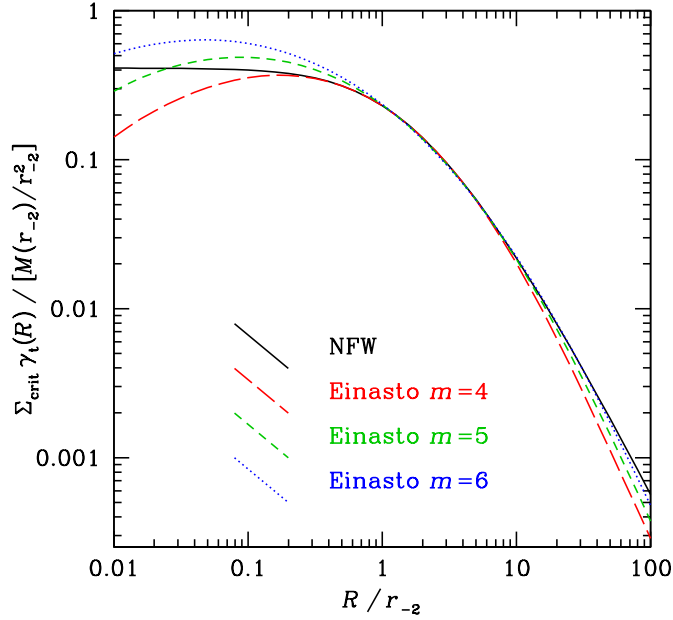}
\caption{Dimensionless tangential shear profile for the NFW model (\emph{black}) and the
  $m=4$ (\emph{red, long-dashed}), 5 (\emph{green, short-dashed}) and 6
  (\emph{blue, dotted}) Einasto models,
  using equation~(\ref{sheardef}) with equations~(\ref{Sigmadef}), (\ref{Rapx}), (\ref{Sigmaapx}),
 (\ref{muapx}), and (\ref{Mprojapx}).}
\label{shear}
\end{figure}
Figure~\ref{shear} shows the subtle differences in the shear profile between
the NFW and Einasto models of index $m=4$, 5, and 6. 
While the tangential shear of the four models is indistinguishable in the
wide range $0.8 < R/r_{-2} < 10$, there are potentially measurable
differences at $R > 10\,r_{-2}$ (at $100\,r_{-2}$, the NFW shear is 1.5 times greater than that for
the $m$=5 Einasto model)  and possibly at $R < 0.8 \,r_{-2}$ (as long
as the weak linear approximation assumed for the measured shear to match the
expression of 
$\gamma_{\rm t}$
of eq.~[\ref{sheardef}] remains valid). 

\section{Surface density and projected mass of the NFW model with lines of sight
  limited to a sphere}

\label{appnfwsph}
In this appendix, we derive the surface density and projected mass (or,
equivalently, projected number) profiles
of the NFW model, with the lines of sight restricted to a sphere (which we
conveniently choose as the virial sphere)
instead of extending to infinity.

\subsection{Surface density profile}

In an analogous manner as the case with line-of-sight extending to infinity
(eq.~[\ref{projec}]), 
the surface density at projected radius $R$ within the sphere of radius
$r_{\rm max}$ is
\begin{equation}
\Sigma^{\rm sph}(R;r_{\rm max}) = 2\,\int_R^{r_{\rm max}} \rho(r) {r\,\d r\over \sqrt{r^2-R^2}} \ .
\label{sbrsphdef}
\end{equation}
We now consider the case of the virial sphere: $r_{\rm max} = r_{\rm v}$.
The surface density can
then
be
written
\begin{equation}
\Sigma^{\rm sph}(R;r_{\rm v}) = {M\left(r_{-2}\right)\over \pi r_{-2}^2}
\,
\widetilde \Sigma^{\rm sph} \left ({R\over r_{-2}},{r_{\rm v}\over
  r_{-2}}\right) \ ,
\end{equation}
where
\begin{eqnarray}
\widetilde \Sigma^{\rm sph}(X,c) &=&
{1\over 2\,\ln2-1}\,
\int_{X}^c {\d x\over (1+x)^2\,\sqrt{x^2-X^2}}
\label{sbrhat}\\
&=&  
{1\over 2 \,\ln2-1}\,
\left \{
\begin{array}{ll}
\displaystyle 
{1\over (1-X^2)^{3/2}}\,
\cosh ^{-1}\left[{c+X^2\over (c+1)\, X}\right]
-{1\over (c+1)}\,\frac{\sqrt{c^2-X^2}}{1-X^2}
   & 
\qquad 0<X<1 \ ,\\
& \\
\displaystyle 
\frac{\sqrt{c^2-1} (c+2)}{3 (c+1)^2}+\frac{\left(-2 c^3-4 c^2-c+2\right)
  (X-1)}{5 (c+1)^2 \sqrt{c^2-1}} 
& \qquad X = 1 < c \ ,\\
& \\
\displaystyle 
{1\over (c+1)}\,
\frac{\sqrt{c^2-X^2}}{X^2-1}-
{1\over (X^2-1)^{3/2}}\,
\cos^{-1}\left[{c+X^2\over (c+1)\, X}\right]
& \qquad 1 < X < c \ ,\\
& \\
0 & \qquad X = 0 \hbox{ or } X > c \ ,
\end{array}
\right.
\label{sbrnfwsph}
\end{eqnarray}
where equation~(\ref{sbrhat}) is found by inserting the NFW density profile (eq.~[\ref{rhoNFW}]) into
equation~(\ref{sbrsphdef}).

\subsection{Projected mass profile}

For the NFW model, the projected mass within the virial sphere is
\begin{equation}
M_{\rm p}^{\rm sph}(R;r_{\rm v}) = \int_0^R 2 \,\pi\,S\,\Sigma^{\rm
  sph}(S;r_{\rm v})\,{\rm d}S
= M(r_{-2})\,\widetilde M_{\rm p}^{\rm sph} \left ({R\over r_{-2}},{r_{\rm v}\over
  r_{-2}}\right) \ ,
\end{equation} 
where
\begin{eqnarray} 
\widetilde M_{\rm p}^{\rm sph}(X,c) &=& 2\,\int_0^X Y\,\widetilde\Sigma^{\rm
  sph}(Y,c)\,\d Y
\label{Mphat} \\
&=& {1 \over \ln 2-1/2}\,
\left \{
\begin{array}{ll}
0 & \qquad X=0 \ ,
\\
& \\
\displaystyle 
\frac{\sqrt{c^2\!-\!X^2}-c}{c+1}+\ln \left[\frac{(c+1)
    \left(c-\sqrt{c^2\!-\!X^2}\right)}{X}\right]+{1\over \sqrt{1-X^2}}\,
\cosh ^{-1}\left[\frac{c+X^2}{(c+1) X}\right] &
\qquad    0<X<1 \hbox{ and } X<c \ ,\\
& \\
\displaystyle 
 \frac{\sqrt{c^2\!-\!X^2}-c}{c+1}+\ln \left[\frac{(c+1)
   \left(c-\sqrt{c^2\!-\!X^2}\right)}{X}\right]+
{1\over \sqrt{X^2-1}}\,
\cos^{-1}\left[\frac{c+X^2}{(c+1) X}\right]
&\qquad  1<X<c \ ,\\
& \\
\displaystyle 
 \ln \left[(c+1) \left(c-\sqrt{c^2-1}\right)\right]-\frac{c}{c+1}+2
 \sqrt{\frac{c-1}{c+1}} 
&\qquad 1=X<c  \ , \\
& \\
\displaystyle 
\ln (c+1)-\frac{c}{c+1}
&\qquad  X\geq c \ ,
\end{array}
\right.
\label{Mprojnfwsph}
\end{eqnarray}
where equation~(\ref{Mprojnfwsph}) was found by inserting
equation~(\ref{sbrnfwsph}) into equation~(\ref{Mphat}).
For $X\geq c$, one recovers the mass within the virial sphere.

\section{Maximum likelihood estimates}
\label{appmle}
In this appendix, we illustrate the maximum likelihood calculations that we
have performed.

Given parameters $\vec \theta$, and data points $\bf x$ the MLE is found by minimizing
\begin{equation}
-\ln {\cal L} = -\sum_j \ln p(x_j|\vec \theta) \ ,
\end{equation}
where ${\cal L} = {\displaystyle \prod_j} p(x_j|\vec \theta)$ is the likelihood.

\subsection{Density profile}
The probability of measuring an object (galaxy or dark matter particle) 
at radius $r$ in a spherical model of
concentration $c$ is
\begin{equation}
p(r_j|c) = {4 \pi r^2\, \left[ \nu(r_j;c) + b\right] \over N(r_{\rm max};c) - N(r_{\rm
    min};c) + 4\,\pi\,b\,\left(r_{\rm max}^3-r_{\rm min}^3\right )/3} \ ,
\end{equation}
where $\nu(R)$ and $N(R)$ are respectively the density and number
(proportional to mass)
profiles, 
$r_{\rm min}$ and $r_{\rm max}$ are
respectively the minimum and maximum radii, $c$ is the concentration, while
$b$ is the constant density background.

\subsection{Surface density profile}
The probability of measuring a galaxy at projected radius $R$ in a spherical
model of concentration $c$ and background $b$ is
\begin{equation}
p(R_j|c,b) = {2 \pi R_j\, \left [\Sigma(R_j;c) + \Sigma_{\rm bg} \right]
  \over N_{\rm p}(R_{\rm max};c) - N_{\rm p}(R_{\rm
    min};c) + \pi\,\Sigma_{\rm bg}\,\left (R_{\rm max}^2-R_{\rm min}^2\right)} \ ,
\end{equation}
where $\Sigma(R)$ and $N_{\rm p}(R)$ are respectively the surface density and
projected number (proportional to projected mass)
profiles, $R_{\rm min}$ and $R_{\rm max}$ are
respectively the minimum and maximum projected radii,
$c$ is the concentration, while $\Sigma_{\rm bg}$ is the constant surface density
background.

For the surface density profile $\Sigma(R)$ and the projected number (mass) profile
$N_{\rm p}(R)$,
we use the formulae of 
\cite{LM01} and of appendix~\ref{appsdensEinasto} for the NFW and Einasto
models, respectively.

\subsection{Distribution of interloper velocities}

According to equation~(\ref{gfit}), 
the distribution of interloper line-of-sight \emph{absolute} velocities, $v_j
\equiv |v_{{\rm los},j}|$, is to first order the sum of a
gaussian and a constant term:
\begin{equation}
p(v_j|\sigma_{\rm i},A,B) = {A\,\exp\left[-v_j^2/\left( 2 \sigma_{\rm
      i}^2\right) \right] + B 
\over \sqrt{\pi/2}\,A\,\sigma_{\rm i}\,{\rm erf} \left[\hat\kappa/(\sigma_{\rm
  i}\sqrt{2})\right] +\hat\kappa\,B} \ ,
\label{pofv}
\end{equation}
where the denominator is found by ensuring $\int_0^{\hat\kappa} p(v_j)\,\d v_j
= 1$ (eq.~[\ref{SigmaivsA}]), and
where $\hat\kappa$ is the maximum considered value of $|v_{\rm los}|/v_{\rm
  v}$ (so $\hat\kappa=4$ in 
Figs.~\ref{vhists}).
If $A$ and $B$ are expressed in virial units, then the denominator of
equation~(\ref{pofv}) is the surface density of particles under consideration in
virial units, which we directly measure from the simulation as $\Sigma=(N/N_{\rm v})/S$,
where $N_{\rm v}$ is the number of particles within the virial sphere, while
$N$ is the number of particles in the radial bin (or within the full virial
cone), and $S$ is the surface of the radial bin (i.e. $\pi$ for the full virial cone).
Hence, substituting for $A=\Sigma\,(1-\hat\kappa\,B')/[\sqrt{\pi/2}\sigma_{\rm i}\,{\rm
    erf}[\hat\kappa/(\sigma_{\rm i}\sqrt{2})]$, we can write the probability of
  measuring an 
interloper absolute velocity as
\begin{equation}
p(v_j|\sigma_{\rm i},B) = {\left (1-\hat\kappa\,B'\right)\,\exp\left[-v_j^2/\left( 2
  \sigma_{\rm i}^2\right) \right]\over \sqrt{\pi/2}\,\sigma_{\rm i}\,{\rm
    erf}\left[\hat\kappa/(\sigma_{\rm i}\sqrt{2})\right]} + B'
\ ,
\label{pofv2}
\end{equation}
where $B'=B/\Sigma$.
Then given the respective uncertainties $\epsilon(\sigma_{\rm i})$ and
$\epsilon(B')$ in $\sigma_{\rm i}$ and $B'$, we deduce the uncertainties in $B$ and
$A$ as
\begin{eqnarray}
\epsilon(B) &=& \Sigma\,\epsilon(B') \ , \\
\epsilon(A) &=& \sqrt{
\left ({\partial A\over \partial \sigma_{\rm
    i}}\right)^2\,\epsilon^2(\sigma_{\rm i}) 
+ \left ({\partial A\over \partial B'}\right)^2\,\epsilon^2(B')
} =
 \Sigma\,{\sqrt{2 \hat\kappa^2\,\pi\,\sigma_{\rm i}^4 {\cal E}^2\,\epsilon^2(B') +
     (1-\hat\kappa\,B')^2\,
\left \{\sqrt{2\pi}\,\sigma_{\rm i}\,{\cal E}-2\hat\kappa\,\exp\left[-\hat\kappa^2/(2\sigma_{\rm
        i}^2)\right]\right\}^2\,\epsilon^2(\sigma_{\rm i})}
\over \pi\,\sigma_{\rm i}^3\,{\cal E}^2
} \ ,
\end{eqnarray}
where ${\cal E} = {\rm erf}\left[\hat\kappa/(\sigma_{\rm i}\sqrt{2})\right]$.

\subsection{Practical considerations}

For one-parameter fits, we first search on a wide linear grid of
equally-spaced 11 points for $\theta_j$, then we
consider the three points with the lowest values of $\-\ln {\cal L}$ and
create a subgrid of 11 equally-spaced points (thus typically zooming in by a
factor of 5), and iterate with finer subgrids
until the two values of the parameter $\theta_j$
with the highest likelihoods differ
 by less than $0.0001$ or when the lowest $-\ln {\cal L}$ decreases by less
 than $10^{-12}$. We then obtain the $1\,\sigma$
confidence interval fitting a cubic spline to the points
below and above the best-fit parameter to solve for $-\ln {\cal L} = - \ln
{\cal L}_{\rm ML} + 0.5$. 

For two-parameter (three-parameter) fits, we first search on a wide 
rectangular (cuboidal) grid of
equally-spaced 11 points. Then we consider the rectangle (cuboid) obtained by
searching for the lowest values of $-\ln {\cal L}$, such that there are at
least 3 different values for both (all three) parameters. We create a sub-grid in this
rectangle (cuboid) with again $11\times11$ ($11\times11\times11$)
points, 
and iterate with finer subgrids until the pair of each of the two (three) parameters with
the highest two likelihoods differ 
by less than $0.0001$ or when the lowest $-\ln {\cal L}$ decreases by less
 than $10^{-12}$.
We then obtain the $1\,\sigma$ contour by considering those points in
parameter space for which  $-\ln {\cal L} = - \ln
{\cal L}_{\rm ML} + 1.15$ (1.77), and then define as the minimum and maximum values
  for each parameter the extreme values in this contour.

\end{document}